\pdfoutput=1

\documentclass[aip,jcp,reprint,groupedaddress,footinbib]{revtex4-1}
\usepackage{graphicx} %
\usepackage{amsmath}
\usepackage{mathtools} %
\usepackage{bm}
\usepackage{bbold}

\usepackage{dcolumn} %
\newcolumntype{d}[1]{D{.}{.}{#1}} %

\usepackage{hyperref}
\hypersetup{colorlinks=true,linkcolor=blue,citecolor=blue,urlcolor=blue}

\usepackage{braket} %

\newcommand{\eu}{\mathrm{e}^}

\newcommand{\rmd}{\mathrm{d}}
\newcommand{\half}{{\ensuremath{\tfrac{1}{2}}}}

\providecommand{\mat}[1]{\mathsf{#1}}
\renewcommand{\mathbf}[1]{\bm{#1}}
\newcommand{\del}{\mathbf{\nabla}}

\DeclareMathOperator{\Imag}{Im}

\newcommand{\ImF}{\mbox{$\Imag F$}}

\newcommand{\der}[3][]{\frac{\rmd^{#1}{#2}}{\rmd{#3}^{#1}}}
\newcommand{\pder}[3][]{\frac{\partial^{#1}{#2}}{\partial{#3}^{#1}}}

\newcommand{\eqn}[1]{Eq.\,(\ref{#1})}

\newcommand{\eqs}[1]{Eqs.\,(\ref{#1})}
\newcommand{\fig}[1]{Fig.~\ref{fig:#1}}
\newcommand{\secref}[1]{Sec.~\ref{sec:#1}\@}
\newcommand{\Ref}[1]{Ref.~\onlinecite{#1}}

\newcommand{\head}[1]{\multicolumn{1}{c}{#1}}

\begin{document}

\title{Ring-polymer instanton theory of electron transfer in the nonadiabatic limit}

\author{Jeremy O. Richardson}
\email{jeremy.richardson@fau.de}
\affiliation{
Institut f{\"u}r Theoretische Physik und Interdisziplin{\"a}res Zentrum f{\"u}r Molekulare Materialien,
Friedrich-Alexander-Universit{\"a}t Erlangen-N{\"u}rnberg (FAU),
Staudtstra{\ss}e 7/B2,
91058 Erlangen, Germany
}

\date{\today}

\begin{abstract}
We take the golden-rule instanton method derived in the previous paper
[arXiv:1509.04919]
and reformulate it using a ring-polymer approach.
This gives equations which can be used to
compute the rates of electron-transfer reactions in the nonadiabatic (golden-rule) limit
numerically
within a semiclassical approximation.
The multidimensional ring-polymer instanton trajectories are obtained efficiently by minimization of the action.
In this form, comparison with Wolynes' quantum instanton method
[P. G. Wolynes, J. Chem. Phys. \textbf{87}, 6559 (1987)]
is possible
and we show that our semiclassical approach is the steepest-descent limit of this method.
We discuss advantages and disadvantages of both methods
and give examples of where the new approach is more accurate.
\end{abstract}

\maketitle

\section{Introduction}

In the previous paper, henceforth referred to as Paper I, \cite{GoldenGreens}
we outlined a derivation of a golden-rule instanton theory
for computing electron-transfer rates in the nonadiabatic limit.
This was based on a time-independent methodology using Fermi's golden rule,
which is correct in the limit that the electronic coupling is weak.
In these equations, we substituted the semiclassical limit of the Green's functions
describing nuclear dynamics on one of two potential-energy surfaces at a given energy.
A number of steepest-descent integrations led to a formula
which defines the rate in terms of the action of an imaginary-time periodic orbit, known as the instanton.

In this paper, we show how this approximate formulation of the rate can be evaluated numerically
to treat electron transfer in large complex systems.
We describe how the golden-rule instanton trajectory can be discretized,
allowing it to be located efficiently using multidimensional optimization techniques.
This is done using a ring-polymer instanton approach
similar to that used by related methods employing a single Born-Oppenheimer surface,
including
the adiabatic rate
\cite{rpinst,Andersson2009Hmethane,Andersson2011HCO,
	Goumans2010Hbenzene,*Goumans2011Hmethanol,*Meisner2011isotope,*Rommel2012enzyme,
	Rommel2011locating,Rommel2011grids,
	Althorpe2011ImF,DMuH}
as well as tunnelling splitting calculations.
\cite{tunnel,water,octamer,Kawatsu2014RPI,*Kawatsu2015NH3}

In contrast, early applications of instanton approaches
employed a method known as ``shooting'' to locate the required instanton trajectory.
This method ran classical dynamics on the inverted potential-energy surface
and attempted to choose 
the correct initial conditions such that the trajectory closed into a periodic orbit. \cite{Chapman1975rates}
Because the trajectories are unstable,
this approach is inefficient and in general limited to treating systems of very few dimensions. \cite{Benderskii}

Many alternative methods exist 
for computing nonadiabatic rate constants
based on a time-dependent formulation.
These include
exact wave function calculations
\cite{Wang2001hybrid,*Thoss2001hybrid,Wang2006flux}
and real-time path-integral calculations
for system-bath models.
\cite{Marchi1991tunnelling,
	Topaler1996nonadiabatic,
	Muehlbacher2003spinboson,*Muehlbacher2004asymmetric,Ananth2012QUAPI}
For more general systems approximate trajectory-based methods have been developed
\cite{Huo2013PLDM,*Huo2015PLDM,Kapral2015QCL,
	Sun1998mapping,*Wang1999mapping,Cotton2014ET,
	Schwerdtfeger2014ET,Landry2012hopping}
including extensions of ring-polymer molecular dynamics.
\cite{mapping,Ananth2013MVRPMD,
	Menzeleev2011ET,*Kretchmer2013ET,*Menzeleev2014kinetic,
	Shushkov2012RPSH}

There are some difficulties with time-dependent methods however,
as the flux correlation functions \cite{Miller1983rate} can become very oscillatory
when describing electron transfer. \cite{Huo2013PLDM}
Some work towards avoiding these problems has been achieved by
modifying the correlation function formalism to remove the oscillations,
although without affecting the long-time limit which defines the exact rate. \cite{nonoscillatory}
This simplification was achieved in part by considering a time-independent picture,
as we have also done in the derivation of the golden-rule instanton.

Although the derivation is very different,
we also show how our result can be related
to Wolynes' quantum instanton method. \cite{Wolynes1987nonadiabatic}
This approach uses an approximation based on the short-time behaviour of the flux correlation function in the nonadiabatic (golden-rule) limit
and is evaluated using path-integral Monte Carlo.
The method has been applied
to study electron transfers in chemically and biologically relevant systems.
\cite{Bader1990golden,Zheng1989ET,*Zheng1991ET}
Our new derivation of a golden-rule rate
offers more insight into the approximations made by such methods
and is in some cases more accurate.

An outline of the paper is as follows.
The main results from Paper I are summarized in \secref{instanton},
and we show how the action integral is discretized and its derivatives obtained in \secref{derivatives}\@.
We thus obtain a ring-polymer instanton formulation for the electron-transfer rate,
which is related to Wolynes' quantum instanton approach in \secref{RPI}\@.
Suggestions for how the instanton approach could be applied numerically to complex systems
are presented in \secref{numerical},
which introduces an efficient algorithm for locating the instanton trajectories.
This is applied to an example system in \secref{results}\@ to analyse its convergence properties,
and \secref{conclusions} concludes the article.

\section{Summary of the Golden-Rule Instanton Approach}
\label{sec:instanton}

It was shown in Paper I
that the instanton relevant to the electron-transfer problem
is an imaginary-time periodic orbit.
This is formed of two trajectories which
travel on the upside-down reactant, $V_0(\mat{x})$, or product, $V_1(\mat{x})$, $f$-dimensional potential-energy surfaces.
They each bounce once and join smoothly together at a point, $\mat{x}^\ddag$, found on the crossing seam, defined by $V_0(\mat{x})=V_1(\mat{x})$.

In this paper, we deal only with imaginary-time trajectories
and thus depart from the notation of Paper I by dropping the bar over imaginary properties.
The Euclidean action along one trajectory,
either on the reactant $(n=0)$ or product $(n=1)$ surface,
is \cite{Feynman,Miller1971density}
\begin{align}
	S_n \equiv
	S_n(\mat{x}',\mat{x}'',\tau_n)
	&= \int_0^{\tau_n} \left[\half m \left|\pder{\mat{x}(\tau)}{\tau}\right|^2 + V_n\big(\mat{x}(\tau)\big) \right] \rmd \tau,
	\label{Sbar}
\end{align}
where %
the trajectory,
$\mat{x}(\tau)$,
travels through the classically forbidden region from $\mat{x}(0)=\mat{x}'$ to $\mat{x}(\tau_n)=\mat{x}''$,
or equivalently in the opposite direction.
A complete periodic orbit which runs in imaginary time $\beta\hbar$
has the action
\begin{align}
	{S}(\mat{x}',\mat{x}'',\tau) = {S}_0(\mat{x}',\mat{x}'',\beta\hbar-\tau) + {S}_1(\mat{x}'',\mat{x}',\tau),
	\label{S}
\end{align}
where $\tau\in[0,\beta\hbar]$.
The particular periodic orbit required is that which
is stationary in $\mat{x}'$, $\mat{x}''$ and $\tau$.
In the following all terms are evaluated at this stationary point,
at which $\mat{x}'=\mat{x}''=\mat{x}^\ddag$.

The golden-rule instanton method derived in Paper I
gives a semiclassical approximation to the rate
in terms of the actions along these trajectories.
Two equivalent formulae are
\begin{align}
	k_\text{SC} Z_0
	&=
			\sqrt{2\pi\hbar} \, \frac{\Delta^2}{\hbar^2}
			\sqrt\frac{{C}_0 {C}_1}{-\Sigma}
			\, \eu{-{S}/\hbar}
	\label{kSpartial}
\\
	&=
			\sqrt{2\pi\hbar} \, \frac{\Delta^2}{\hbar^2}
			\sqrt{\frac{{C}_0{C}_1}{{C}}}
			\left(-\frac{\rmd^2 {S}}{\rmd\tau^2}\right)^{-\half} 
			\,\eu{-{S}/\hbar},
	\label{kSfull}
\end{align}
where the van-Vleck prefactor for a trajectory is given by
\begin{align}
	{C}_n &= \left|-\frac{\partial^2{S}_n}{\partial\mat{x}'\partial\mat{x}''}\right|
\end{align}
and the other prefactors are
\begin{align}
		{C} &= \left| \begin{matrix} \frac{\partial^2 {S}}{\partial\mat{x}'\partial\mat{x}'} &
										 \frac{\partial^2 {S}}{\partial\mat{x}'\partial\mat{x}''} \\
										 \frac{\partial^2 {S}}{\partial\mat{x}''\partial\mat{x}'} &
										 \frac{\partial^2 {S}}{\partial\mat{x}''\partial\mat{x}''}
						 \end{matrix} \right|
\\
		\Sigma &= \left| \begin{matrix} \frac{\partial^2 {S}}{\partial\mat{x}'\partial\mat{x}'} &
										 \frac{\partial^2 {S}}{\partial\mat{x}'\partial\mat{x}''} &
										 \frac{\partial^2 {S}}{\partial\mat{x}' \partial\tau} \\
										 \frac{\partial^2 {S}}{\partial\mat{x}''\partial\mat{x}'} &
										 \frac{\partial^2 {S}}{\partial\mat{x}''\partial\mat{x}''} &
										 \frac{\partial^2 {S}}{\partial\mat{x}'' \partial\tau} \\
										 \frac{\partial^2 {S}}{\partial\tau \partial\mat{x}'} &
										 \frac{\partial^2 {S}}{\partial\tau \partial\mat{x}''} &
										 \frac{\partial^2 {S}}{\partial\tau^2} \\
						 \end{matrix} \right|.
		\label{Sigma}
\end{align}

These formulae for the golden-rule instanton method
were used in Paper I to obtain the rate of electron transfer
in a few special systems for which the bounce trajectories and corresponding action is known analytically.
In order to apply the method to more general problems with anharmonic potentials,
we will require numerical methods
which are able to locate the instanton trajectory
and evaluate the action and its derivatives.
This is the topic addressed in this paper.

\section{Discretization Scheme}
\label{sec:derivatives}

In this section, %
we show how the
action integral, \eqn{Sbar}, can be defined
from a discretized form of an imaginary-time trajectory.
This is based on the ring-polymer instanton method, \cite{rpinst}
which has been successfully used in adiabatic, single-surface, rate calculations \cite{Andersson2009Hmethane,Rommel2011locating}
as well as the evaluation of tunnelling splittings.  \cite{tunnel,water,octamer}
It relies on the fact that a classical trajectory
is known to give a stationary value of the action,
with respect to any deviation along its length except at the end points. \cite{Whittaker,Goldstein}

We also describe how second derivatives of the action can be evaluated directly
without resorting to taking finite differences between instantons optimized under various conditions. %
The approach we use for this
follows closely the method of implicit differentiation described in \Ref{Althorpe2011ImF},
which we extend to obtain all the derivatives required for the golden-rule instanton method.

According to our golden-rule approach, \cite{GoldenGreens}
we only need to study the dynamics on one of
the two potential-energy surfaces at any time.
This section would thus
also be directly applicable to single-surface reactions,
simply by dropping the subscript $n$. %

We consider an imaginary-time pathway of length $\tau_n$ between the points $\mat{x}'\equiv\mat{x}_0$ and $\mat{x}''\equiv\mat{x}_{N_n}$,
which passes through the intermediate points $\{\mat{x}_1,\dots,\mat{x}_{N_n-1}\}$
at a set of discrete times.
The imaginary-time intervals between each point are $\delta\tau_i=\epsilon_i\tau_n$,
with $i\in\{1,\dots,N_n\}$
such that each $\epsilon_i\in[0,1]$ and $\sum_{i=1}^{N_n} \epsilon_i=1$.
The velocity along a given pathway at these times is given 
by $|\mat{x}_i-\mat{x}_{i-1}|/\epsilon_i\tau_n$
and the action by
\begin{multline}
	\label{SN}
	S_n(\mat{x}_0,\dots,\mat{x}_{N_n};\tau_n) =
		\sum_{i=1}^{N_n} \frac{m|\mat{x}_i - \mat{x}_{i-1}|^2}{2\epsilon_i\tau_n}
		\\
		+ \sum_{i=1}^{N_n} \epsilon_i\tau_n \frac{V_n(\mat{x}_{i-1}) + V_n(\mat{x}_i)}{2},
\end{multline}
where the first term originates from a trapezium-rule integration of the kinetic energy along the pathway,
and the second of the potential energy.
This is the general form allowing for uneven imaginary time intervals \cite{Rommel2011grids}
and would simplify to the usual case with $\epsilon_i=1/{N_n}$.
Note that here an open-ended pathway is described such that no cyclic indices are implied.

The points $\tilde{\mat{x}}_i$ for $i\in\{1,\dots,N_n-1\}$,
which give the coordinates along the classical trajectory, 
are those which give a stationary value of $S_n$,
i.e.\ those which solve
\begin{align} \label{dSdx}
		\frac{\tilde{\mat{x}}_i - \tilde{\mat{x}}_{i-1}}{\epsilon_i}
		+ \frac{\tilde{\mat{x}}_i - \tilde{\mat{x}}_{i+1}}{\epsilon_{i+1}}
		+ \frac{(\epsilon_i+\epsilon_{i+1})\tau_n^2}{2m} \pder{V_n}{\tilde{\mat{x}}_i}
		= \mat{0}.
\end{align}
The action along the trajectory is therefore
$S_n(\mat{x}',\mat{x}'',\tau_n) \equiv \lim_{N_n\rightarrow\infty} \tilde{S}_n(\mat{x}',\mat{x}'',\tau_n)$,
where
$\tilde{S}_n(\mat{x}',\mat{x}'',\tau_n) = S_n(\tilde{\mat{x}}_0,\tilde{\mat{x}}_1,\dots,\tilde{\mat{x}}_{N_n};\tau_n)$
and
$\tilde{\mat{x}}_0\equiv\mat{x}'$, $\tilde{\mat{x}}_{N_n}\equiv\mat{x}''$.

In fact the dominant classical trajectory between two end points in a given imaginary time 
will be the global minimum of \eqn{SN} with respect to the intermediate points.
This can be obtained by employing a multidimensional optimization routine
such as the limited memory Broyden-Fletcher-Goldfarb-Shanno (l-BFGS) algorithm \cite{LBFGSB1997algorithm}
in the same way as is done for tunnelling splitting calculations. \cite{tunnel,water}
However, the end points, $\mat{x}^\ddag$, required for the instanton method
are not in general known \textit{a priori},
so the instanton trajectories cannot be obtained in this way.
We discuss optimization methods which do not require knowledge of the end points in \secref{RPI}.

Differentiating \eqn{dSdx} by the end points, $\mat{x}'$ or $\mat{x}''$, gives
equations which can be written in the following form:
\begin{subequations}
\label{Jdx}
\begin{align}
	\sum_{i'=1}^{N_n-1} \sum_{j'=1}^{f} J_{ij,i'j'} \pder{\tilde{x}_{i'j'}}{x'_k} &= \frac{\delta_{i1}\delta_{jk}}{N_n \epsilon_1}
\\
	\sum_{i'=1}^{{N_n}-1} \sum_{j'=1}^{f} J_{ij,i'j'} \pder{\tilde{x}_{i'j'}}{x''_k} &= \frac{\delta_{i\,{N_n}-1}\delta_{jk}}{N_n \epsilon_{N_n}},
\end{align}
\end{subequations}
where
the elements of the doubly indexed matrix $\mathbf{J}$ are defined by
\begin{multline}
	J_{ij,i'j'} = 
		\frac{\delta_{ii'} - \delta_{i+1\, i'}}{N_n\epsilon_i} \delta_{jj'}
		+ \frac{\delta_{ii'} - \delta_{i-1\,i'}}{N_n\epsilon_{i+1}} \delta_{jj'}
		\\
		+ \delta_{ii'}\frac{(\epsilon_i+\epsilon_{i+1})\tau_n^2}{2N_nm} \nabla^2_{jj'} V_n(\tilde{\mat{x}}_i). 
	\label{J}
\end{multline}
Again the indices are not cyclic,
i.e.\ the matrix is banded
with bandwidth $f$.
This definition is equivalent to that given in \Ref{Althorpe2011ImF} when the time-steps are equal.
Equations \eqref{Jdx} can be solved numerically for the derivatives of $\tilde{\mat{x}}$
using standard linear-algebra routines.
Note that these partial derivatives imply that $\tau_n$ and one end point are kept fixed
while the rest of the pathway is allowed to re-optimize itself as the other end point varies.

Other terms are found by differentiating \eqn{dSdx} by $\tau_n$ to give
\begin{align}
	\label{Jdt}
	\sum_{i'=1}^{N_n-1} \sum_{j'=1}^{f} J_{ij,i'j'} \pder{\tilde{x}_{i'j'}}{\tau_n} &= -\frac{(\epsilon_i+\epsilon_{i+1})\tau_n}{N_n m} \pder{V_n}{\tilde{x}_{ij}},
\end{align}
which we solve for $\pder{\tilde{\mat{x}}_i}{\tau_n}$.

Using the fact that $\tilde{S}_n$ is stationary with respect to differentiation by $\tilde{\mat{x}}_i$
gives
\begin{subequations}
\begin{align}
	\frac{\partial \tilde{S}_n}{\partial \mat{x}'}
		&= \frac{m (\mat{x}' - \tilde{\mat{x}}_1) }{\epsilon_1\tau_n} + \frac{\epsilon_1\tau_n}{2} \del V_n(\mat{x}')
\\
	\frac{\partial \tilde{S}_n}{\partial \mat{x}''}
		&= \frac{m (\mat{x}'' - \tilde{\mat{x}}_{N_n-1})}{\epsilon_{N_n}\tau_n} + \frac{\epsilon_{N_n}\tau_n}{2} \del V_n(\mat{x}'') 
\\
	\frac{\partial \tilde{S}_n}{\partial \tau_n}
		&=- \sum_{i=1}^{N_n} \frac{m|\tilde{\mat{x}}_i - \tilde{\mat{x}}_{i-1}|^2}{2\epsilon_i\tau_n^2}
		+ \sum_{i=1}^{N_n} \epsilon_i \frac{V_n(\tilde{\mat{x}}_{i-1}) + V_n(\tilde{\mat{x}}_i)}{2}.
\end{align}
\end{subequations}
Differentiating again,
we obtain the second derivatives:
\begin{subequations}
\begin{align}
	\frac{\partial^2 \tilde{S}_n}{\partial x_j' \partial x_k'}
		&= \frac{m}{\epsilon_1\tau_n}\left(\delta_{jk} - \pder{\tilde{x}_{1j}}{x'_k}\right) + \frac{\epsilon_1\tau_n}{2} \nabla^2_{jk} V_n(\mat{x}') \\
	\frac{\partial^2 \tilde{S}_n}{\partial x_j' \partial x_k''}
		&= - \frac{m}{\epsilon_1\tau_n} \pder{\tilde{x}_{1j}}{x''_k} \\
	\frac{\partial^2 \tilde{S}_n}{\partial x_j'' \partial x_k''}
		&= \frac{m}{\epsilon_{N_n}\tau_n}\left(\delta_{jk} - \pder{\tilde{x}_{{N_n}-1\,j}}{x''_k}\right) + \frac{\epsilon_{N_n}\tau_n}{2} \nabla^2_{jk} V_n(\mat{x}'') \\
	\frac{\partial^2 \tilde{S}_n}{\partial \mat{x}' \partial \tau_n}
		&= - \frac{m(\mat{x}'-\tilde{\mat{x}}_1)}{\epsilon_1\tau_n^2} - \frac{m}{\epsilon_1\tau_n} \pder{\tilde{\mat{x}}_{1}}{\tau_n} + \frac{\epsilon_1}{2} \pder{V_n}{\mat{x}'} \\
	\frac{\partial^2 \tilde{S}_n}{\partial \mat{x}'' \partial \tau_n}
		&= - \frac{m(\mat{x}''-\tilde{\mat{x}}_{{N_n}-1})}{\epsilon_{N_n}\tau_n^2} - \frac{m}{\epsilon_{N_n}\tau_n} \pder{\tilde{\mat{x}}_{{N_n}-1}}{\tau_n} + \frac{\epsilon_{N_n}}{2} \pder{V_n}{\mat{x}''} \\ 
	\frac{\partial^2 \tilde{S}_n}{\partial \tau_n^2}
		&= \sum_{i=1}^{N_n} \frac{m|\tilde{\mat{x}}_i - \tilde{\mat{x}}_{i-1}|^2}{\epsilon_i\tau_n^3}
			+ \sum_{i=1}^{{N_n}-1} \frac{\epsilon_i+\epsilon_{i+1}}{2} \pder{V_n}{\tilde{\mat{x}}_{i}} \cdot \pder{\tilde{\mat{x}}_i}{\tau_n}
		\nonumber \\ &\quad
			- \sum_{i=1}^{{N_n}-1} \frac{m}{\tau_n^2} \left(\frac{\tilde{\mat{x}}_i-\tilde{\mat{x}}_{i-1}}{\epsilon_i} + \frac{\tilde{\mat{x}}_i-\tilde{\mat{x}}_{i+1}}{\epsilon_{i+1}}\right) \cdot \pder{\tilde{\mat{x}}_i}{\tau_n}.
\end{align}
\end{subequations}
Partial derivatives of $S_n$
are approximated by these formulae,
which become exact in the ${N_n}\rightarrow\infty$ limit.
Assuming that the instanton trajectories have already been found,
these derivatives can be applied in the prefactor of \eqn{kSpartial}, using \eqn{Sigma} and \eqn{S},
to give the golden-rule instanton rate.

In contrast to standard approaches where the eigenvalues of a $Nf\times Nf$ matrix are required
for the prefactor,
the most difficult task in this approach is the solution of the linear equations, Eqs.~\eqref{Jdx} and \eqref{Jdt}.
Because $\mathbf{J}$ is the Hessian matrix about the minimum pathway, it is positive definite,
and the equations can be solved efficiently using a Cholesky decomposition,
taking advantage of the banded nature of the matrix. \cite{NumRep}

This approach is not limited to the current application
but may also significantly improve the efficiency of other instanton methods,
for which the diagonalization can be a considerably time-consuming task for high-dimensional systems.
We shall discuss the use of such an approach to improve the efficiency
of adiabatic rate calculations in a forthcoming paper.

\section{Ring-polymer instanton formulation}
\label{sec:RPI}

So far, we have only dealt with open-ended trajectories, whose end points are as yet unknown.
In this section, we extend this methodology to obtain the pathway for the total periodic orbit.
This orbit is simply the combination of a trajectory on the reactant surface with another on the product surface
and has total imaginary time $\beta\hbar$.

\begin{figure}
\includegraphics{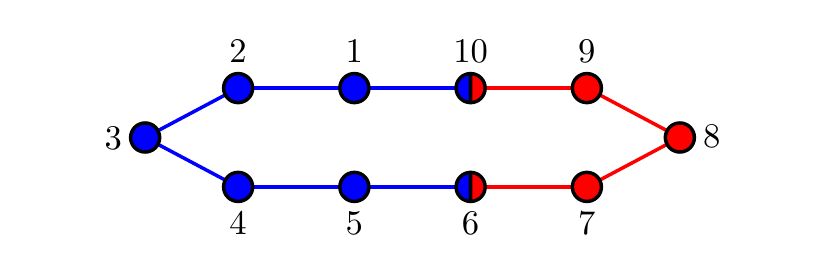}
\caption{Schematic showing the ring-polymer beads which discretize the instanton
for the case of $N_0=6$ and $N_1=4$.
Those on the left coloured in blue have the electronic configuration of the reactant state $\ket{0}$
and those on the right in red of the product $\ket{1}$.
Beads $N_0$ and $N$ are located at the hopping point $\mat{x}^\ddag$ and contribute to the action of both trajectories.
}
\label{fig:beads}
\end{figure}

We divide up the total orbit into $N$ segments,
with the first $N_0$ on $\ket{0}$ and the remaining $N_1=N-N_0$ on $\ket{1}$ as in \fig{beads}.
Equal time-step intervals, $\epsilon_i=1/N_n$, will be assumed here
but other choices may slightly improve efficiency. \cite{Rommel2011grids}

There is a special case that the time intervals on both trajectories are equal, all with length $\beta_N\hbar$,
where $\beta_N = \beta/N$.
This can only be obtained in practice
if the imaginary times along each trajectory, $\tau_n=N_n\beta_N\hbar$, are known \textit{a priori}.
Such cases arise
for example if the reaction is symmetric, where the stationary value is known to be $\tau=\beta\hbar/2$,
or if the instanton has already been obtained by an alternative method, such as those introduced in \secref{numerical}.

In this case, we have a formulation similar to
path-integral \cite{Parrinello1984Fcenter} and ring-polymer molecular dynamics, \cite{rpinst,Habershon2013RPMDreview}
which were obtained from a discretization of the quantum Boltzmann operator.
The resulting set of $N$ coordinates are called beads
and, via a quantum-classical correspondence, \cite{Chandler+Wolynes1981}
are equivalent to a ring polymer of classical particles connected together by harmonic springs.

It is a good idea to use the $N$-bead steepest-descent approximation to the reactant partition function, \cite{Kleinert}
\begin{align}
	Z_0 &= \prod_{j=1}^f \left[2\sinh\frac{\beta\hbar\tilde{\omega}_j}{2}\right]^{-1}
\\
	\tilde{\omega}_j &= \frac{2}{\beta_N\hbar} \sinh^{-1}\frac{\beta_N\hbar\omega_j}{2},
\end{align}
as this is known to benefit from a cancellation of errors with the $N$-bead instanton calculation
and improve convergence of the rate. \cite{Andersson2009Hmethane}
Here $\omega_j$ are the normal-mode frequencies at the minimum of $V_0(\mat{x})$;
if there are translation or rotational modes, the formula should be modified appropriately.

The total action along the two joined pathways
is given by %
\begin{multline}
	\beta_N U_N(\mathbf{x}) = S_0(\mat{x}_N,\mat{x}_1,\dots,\mat{x}_{N_0};N_0\beta_N\hbar)
	\\
	+ S_1(\mat{x}_{N_0},\dots,\mat{x}_N;N_1\beta_N\hbar),
	\label{SRPI}
\end{multline}
such that the $N$-bead ring-polymer potential
is
\begin{multline}
	U_N(\mathbf{x}) =
		\sum_{i=1}^N \frac{m}{2\beta_N^2\hbar^2} |\mat{x}_{i+1} - \mat{x}_{i}|^2
	\\
		+ \half V_0(\mat{x}_{N}) + \sum_{i=1}^{N_0-1} V_0(\mat{x}_i) + \half V_0(\mat{x}_{N_0})
	\\
		+ \half V_1(\mat{x}_{N_0}) + \sum_{i=N_0+1}^{N-1} V_1(\mat{x}_i) + \half V_1(\mat{x}_{N}).
	\label{UN}
\end{multline}
The positions of each bead are given by $\mathbf{x}=\{\mat{x}_1,\dots,\mat{x}_{N}\}$,
and cyclic indices are implied such that $\mat{x}_0\equiv\mat{x}_N$.
This function can be minimized with respect to the positions of all beads
to obtain the coordinates $\tilde{\mathbf{x}}=\{\tilde{\mat{x}}_1,\dots,\tilde{\mat{x}}_N\}$ along both trajectories simultaneously.
The hopping point is then identified as $\mat{x}^\ddag=\tilde{\mat{x}}_{N_0}=\tilde{\mat{x}}_N$
and the action as $\tilde{S}=\beta_N\tilde{U}_N$,
where $\tilde{U}_N=U_N(\tilde{\mathbf{x}})$.
However, $\tau$, or equivalently the ratio $N_0/N_1$, is yet to be specified.
It will therefore be necessary to compute the instanton
for numerous values of $\tau$
to find the stationary point of $\tilde{U}_N$ with respect to $\tau$.

We now introduce the quantum instanton approach of Wolynes. \cite{Wolynes1987nonadiabatic}
This method was derived
using a steepest-descent evaluation of the time integral over the exact flux-flux correlation function
within the golden-rule approximation
and gives
\begin{align}
	\label{Wolynes}
	k_\text{QI} Z_0 &= \sqrt{2\pi\hbar} \, \frac{\Delta^2}{\hbar^2} \left(-\der[2]{\phi}{\tau}\right)^{-\half} \eu{-\phi(\tau)/\hbar}
\\
	\eu{-\phi(\tau)/\hbar} &= \Lambda^{-Nf} %
		\int \eu{-\beta_N U_N(\mathbf{x})} \, \rmd\mathbf{x},
	\label{phi}
\end{align}
where the prefactor is $\Lambda=\sqrt{2\pi\beta_N\hbar^2/m}$.
It is here assumed for simplicity, that the electronic coupling, $\Delta$, is approximately constant,
although the formulation could be generalized without affecting our findings.
In practice the integrals are computed using a discrete path-integral Monte Carlo simulation,
and $\tau=N_1\beta_N\hbar$
is chosen in the range $[0,\beta\hbar]$ such that $\der{\phi}{\tau}=0$.
Taking the second derivative of \eqn{phi} gives
\cite{Cao1997nonadiabatic,*Cao1998erratum}
\begin{align}
	\label{d2phi}
		- \frac{1}{\hbar} \der[2]{\phi}{\tau} \, \eu{-\phi(\tau)/\hbar}
		&=
		\frac{\Lambda^{-Nf}}{\hbar^2} %
		\int V_-(\mat{x}') V_-(\mat{x}'')
		\, \eu{-\beta_N U_N(\mathbf{x})} \, \rmd\mathbf{x},
\end{align}
where %
$V_-(\mat{x})=V_0(\mat{x})-V_1(\mat{x})$.
The second derivative of $\phi$ is negative and thus corresponds to a stationary point which is a maximum along $\tau$.

The derivation is similar in spirit
to that used to obtain the quantum instanton approach for Born-Oppenheimer systems described in \Ref{Vanicek2005QI},
as it also employs a steepest-descent integration along the real-time coordinate of a flux-flux correlation function.
The single-surface quantum instanton approach is however not a semiclassical approximation
in the sense that it gives the correct leading order of $\hbar$.
This is most easily seen from the fact that it does not reproduce correct results for a free-particle or in the classical limit. \cite{Miller2003QI}
Wolynes' formula, \eqn{Wolynes}, is also not exact in the high-temperature limit in general.
However, it is known that it reproduces the stationary-phase approximation \cite{Weiss}
for the golden-rule rate of a spin-boson system
and hence also Marcus theory, \cite{Marcus1985review}
which is the correct result for this system in the classical limit.

To show the link between the quantum and semiclassical instanton methods,
we perform a steepest-descent approximation to \eqn{phi} in two steps,
reserving the integrals over beads assigned to $\mat{x}'$ and $\mat{x}''$
until after all others.
This gives
\begin{align}
	\eu{-\phi(\tau)/\hbar} &\approx
		\Lambda^{-2f} %
		\sqrt\frac{1}{|\mathbf{J}_0||\mathbf{J}_1|}
		\nonumber\\&\qquad\times
		\iint_\text{SD}
		\eu{-\tilde{S}_0(\mat{x}',\mat{x}'',\beta\hbar-\tau)/\hbar - \tilde{S}_1(\mat{x}'',\mat{x}',\tau)/\hbar}
		\, \rmd\mat{x}' \rmd\mat{x}''
\\
	&= \sqrt\frac{{C}_0{C}_1}{{C}}
		\, \eu{-\tilde{S}/\hbar},
\end{align}
where $\mathbf{J}_n$ is defined in \eqn{J} with $\epsilon_i=1/N_n$
and
we have used the following result from \Ref{Althorpe2011ImF}:
\begin{align}
	{C}_n &= \left(\frac{m}{\beta_N\hbar}\right)^f |\mathbf{J}_n|^{-1}.
\end{align}
Also, within the steepest-descent approximation, $\der[2]{\phi}{\tau} \approx \der[2]{\tilde{S}}{\tau}$ %
and therefore, taking these semiclassical limits in Wolynes' formula, \eqn{Wolynes}, reproduces the golden-rule instanton rate, \eqn{kSfull}.
This shows a strong link between the semiclassical instanton theory presented in this paper and the quantum instanton approach---%
the former is a steepest-descent approximation to the latter.
The quantum instanton approach has a great advantage over the semiclassical instanton method,
which is that it can also treat liquid systems,
where many minima exist on the ring-polymer potential surface.

Note however
that Wolynes' quantum instanton is not always more accurate than the semiclassical instanton.
In the high-temperature limit, the ring-polymer beads collapse and \eqn{phi} reduces to give
an integral over the centroid mode,
\begin{align}
	\label{highT}
	\lim_{\beta\rightarrow0} \eu{-\phi(\tau)/\hbar} = \left(\frac{m}{2\pi\beta\hbar^2}\right)^{f} \int \eu{-\left[(\beta\hbar-\tau)V_0(\mat{x}) + \tau V_1(\mat{x})\right]/\hbar} \, \rmd\mat{x}.
\end{align}
Using this definition of $\phi(\tau)$ in \eqn{Wolynes}
gives a rate which is not in general equal to that of classical golden-rule transition-state theory. \cite{nonoscillatory}
This is most easily seen from the example of the transfer from a harmonic oscillator to an anharmonic product state,
such as the system discussed in \Ref{nonoscillatory}.
As shown in Paper I, the high-temperature golden-rule instanton rate gives
the exact classical golden-rule transition-state theory limit for this one-dimensional system,
\begin{align}
	k_\text{cl,TST} Z_0
	&= \frac{\Delta^2}{\hbar^2} \sqrt\frac{2\pi m}{\beta} \int \eu{-\beta V_0(x)} \, \delta\big[V_0(x) - V_1(x)\big] \, \rmd x,
\end{align}
whereas \eqn{highT} noticeably does not include a delta function constraining the integral to the crossing seam
and thus gives an incorrect result.

This is at first sight surprising,
as one would naively assume that the steepest-descent approximation reduces the accuracy of the result.
The reason for the discrepancy is that the two methods are based on different approximations.
This example makes it clear that, at least for certain problems,
a more accurate quantum rate theory is obtained from semiclassical considerations
than from Gaussian approximations to the flux correlation function.

The link between the semiclassical and quantum instanton approaches
also suggests that another method could be used to compute the golden-rule instanton rate,
where the steepest-descent integration is taken over all ring-polymer beads simultaneously
giving
\begin{align}
	\eu{-\phi(\tau)/\hbar} &\approx
		\left| \frac{\beta_N^2 \hbar^2}{m} \del^2 \tilde{U}_N \right|^{-\half}
		\, \eu{-\beta_N \tilde{U}_N},
\end{align}
where $\del^2 \tilde{U}_N$ is the Hessian matrix found by differentiating \eqn{UN} by all bead positions $\mat{x}_i$
and is evaluated at the instanton geometry, $\tilde{\mathbf{x}}$.

Because the steepest-descent integrals are evaluated at the hopping point where \mbox{$V_-(\mat{x}^\ddag)=0$},
we have to consider a higher-order term for our semiclassical approximation of \eqn{d2phi}.
This is
\begin{multline}
	- \frac{1}{\hbar} \der[2]{\phi}{\tau} \, \eu{-\phi(\tau)/\hbar}
		\approx
		\frac{\Lambda^{-Nf}}{\hbar^2} %
		\int_\text{SD}
		\left[\pder{V_-(\mat{x}^\ddag)}{\mat{x}^\ddag}\cdot(\mat{x}'-\mat{x}^\ddag)\right]
\\\times
		\left[\pder{V_-(\mat{x}^\ddag)}{\mat{x}^\ddag} \cdot (\mat{x}''-\mat{x}^\ddag) \right]
		\, \eu{-\beta_NU_N(\mathbf{x})}
		\, \rmd\mathbf{x}.
\end{multline}
We evaluate the integral using a second-order expansion of $U_N(\mathbf{x})$ about the ring-polymer instanton orbit,
which gives, %
\begin{align}
	\der[2]{\tilde{S}}{\tau}
		&= - \frac{1}{\hbar} \pder{V_-(\mat{x}^\ddag)}{\mat{x}^\ddag} \cdot [\beta_N\del^2\tilde{U}_N]^{-1}_{N_0,N} \cdot \pder{V_-(\mat{x}^\ddag)}{\mat{x}^\ddag},
\end{align}
where only the $f\times f$ block corresponding to rows for bead $N_0$ and columns for bead $N$ is required from the inverse of the full Hessian.
The golden-rule instanton rate in ring-polymer form is thus
\begin{align}
	\label{kRPI}
	k_\text{SC} Z_0 &= \sqrt{2\pi\hbar} \, \frac{\Delta^2}{\hbar^2}
		\left| \frac{\beta_N^2 \hbar^2}{m} \del^2 \tilde{U}_N \right|^{-\half}
		\left(-\der[2]{\tilde{S}}{\tau}\right)^{-\half} \eu{-\beta_N \tilde{U}_N}.
\end{align}
This formula gives the same result as \eqs{kSpartial} and \eqref{kSfull} in the $N\rightarrow\infty$ limit.

Note that all eigenvalues of the Hessian are positive.
This is therefore a more straightforward derivation
than is achieved using the \ImF\ approach, \cite{rpinst,Cao1997nonadiabatic}
where the instanton has
a negative eigenvalue, which has its sign reversed,
and a zero-mode which has to be integrated out analytically.

As in the adiabatic, single-surface, case, \cite{rpinst}
this ring-polymer instanton approach
provides a computationally tractable way to obtain the reaction rate of a multidimensional system.
However, it would be necessary in general to optimize \eqn{UN} many times
to find the value of $\tau$ which gives a maximum value of $\tilde{U}_N$.
In \secref{numerical}, we shall propose alternative methods which obtain $\tau$ automatically from a single optimization
and may therefore be found to be more efficient
in practical applications.

\section{Numerical Evaluation}
\label{sec:numerical}

In this section we present two methods which we suggest could be used to evaluate semiclassical golden-rule rates
in complex multidimensional systems.
It may also be possible to implement similar schemes for locating other instantons
more efficiently,
including those for adiabatic rate theory \cite{rpinst} and tunnelling splitting calculations. \cite{tunnel,water}
Applications of the methods to such systems will be explored in future work.

In \secref{RPI}, we discussed an approach similar to that used for adiabatic instantons,
where the imaginary time of each trajectory is chosen before the ring-polymer instanton is optimized.
Here we present two alternative methods which optimize all unknown variables simultaneously.
The first is based on a Lagrangian formalism with equal time-steps and the second
uses the Hamilton-Jacobi abbreviated action with evenly spaced ring-polymer beads.

Note that the symmetry of the instanton pathway can be used to reduce the number of independent coordinates to $N/2+1$.
\cite{Andersson2009Hmethane}
It is known that the instanton must follow the same pathway in both directions of its periodic orbit,
such that we only need to optimize two shorter open-ended trajectories, each with one end at the hopping point and the other at a turning point.
In both cases, we will employ the bead ordering given in \fig{beads}
and assume that $N_0$ and $N_1$ are always chosen to be even.
There is a symmetry equivalence between the top and bottom rows
such that when the pathway is optimized, 
$\tilde{\mat{x}}_{N_0/2-i} = \tilde{\mat{x}}_{N_0/2+i}$ for $i\in\{1,\dots,N_0/2\}$ and $\tilde{\mat{x}}_{N_0+N_1/2+i} = \tilde{\mat{x}}_{N_0+N_1/2-i}$ for $i\in\{1,\dots,N_1/2\}$.
The beads at the turning points, $\tilde{\mat{x}}_{N_0/2}$ and $\tilde{\mat{x}}_{N_0+N_1/2}$, are independent.

\subsection{Lagrangian formalism}
\label{sec:event}

The Lagrangian formalism defines classical trajectories according to the elapsed time.
It was used to define the standard ring-polymer instanton approach for single-surface systems with equal time-steps. \cite{rpinst,tunnel,water}
As in \secref{RPI},
we again separate each trajectory into $N_n$ equal imaginary-time intervals,
i.e.\ with $\epsilon_i=1/N_n$.
However in contrast to the previous approach, \eqn{SRPI}, the reactant trajectory may have a different time step from the product trajectory.
The total discretized action is
\begin{multline}
	S(\mathbf{x},\tau) %
	= 2 S_0\big(\mat{x}_{N_0/2},\dots,\mat{x}_{N_0};\half(\beta\hbar-\tau)\big) \\+ 2 S_1\big(\mat{x}_{N_0},\dots,\mat{x}_{N_0+N_1/2};\half\tau\big),
\end{multline}
where due to the forementioned symmetry
we have taken twice the action along each pathway from the turning point to the hopping point in half the imaginary time.
The classical imaginary-time periodic orbit can be found as the first-order saddle point of this function
with respect to the independent bead coordinates $\mathbf{x}=\{\mat{x}_{N_0/2},\dots,\mat{x}_{N_0+N_1/2}\}$ and $\tau$ simultaneously.
The other half of the instanton orbit is given by symmetry.
Saddle-point optimization algorithms have been well studied in the pursuit of locating instantons, \cite{Rommel2011locating,PhD}
in most cases a Hessian-based quasi-Newton method being appropriate.
The value of the optimized function gives the required total action $\tilde{S}$ in the $N$-bead approximation
and the imaginary times $\tau_0=\beta\hbar-\tau$ and $\tau_1=\tau$.
In the $N\rightarrow\infty$ limit, this result is in principle independent of the choice of the ratio $N_0/N_1$,
although an intelligent suggestion would be $\tau_0/N_0\approx\tau_1/N_1$ to make all time-steps approximately equal.

In this way, 
it is possible to evaluate \eqn{kSpartial} numerically
using this ring-polymer instanton approach
and converge the results obtained with respect to $N$.
We identify $\mat{x}'\equiv\tilde{\mat{x}}_{N}$ and $\mat{x}''\equiv\tilde{\mat{x}}_{N_0}$,
both of whose optimized positions
tend to $\mat{x}^\ddag$ in the $N\rightarrow\infty$ limit.
Derivatives of the total action $\tilde{S}$ are
given as sums or differences of the derivatives of $\tilde{S}_0$ and $\tilde{S}_1$
defined in \secref{derivatives}.
Note that here the full trajectory, from $\mat{x}'$ to $\mat{x}''$ is required
and not just the trajectory to the turning point.

However, this approach requires a saddle-point optimization
which is often more difficult than a minimization.
In \secref{HamJac}, we describe an alternative method
to locate the instanton trajectories and evaluate their actions
based on a potentially simpler algorithm.

\subsection{Hamilton-Jacobi formalism}
\label{sec:HamJac}

A significant feature of the derivation presented in this paper
is that the energy of the two trajectories must be equal.
It would therefore be natural
to locate the instanton trajectory under this constraint
rather than directly attempting to find the stationary value of the imaginary time $\tau$.
To this end,
we will employ a Hamilton-Jacobi definition for the action
along two discretized pathways of $N_0$ and $N_1$ ring-polymer beads
with the same energy for each trajectory.

We should take care when computing the discretized abbreviated action,
as a naive implementation using the trapezium rule to approximate $W_n$
would give a function with infinite derivatives at the turning points.
We therefore propose the following functional form
to compute the abbreviated action along one pathway with energy $E$:
\begin{align}
	\label{Wn}
	{W}_n(\mat{x}_0,\dots,\mat{x}_{N_n/2};E) &= 2\sum_{i=1}^{N_n/2} \left|\frac{{p}_n(\mat{x}_{i-1})^3 - {p}_n(\mat{x}_i)^3}{3m\kappa_i}\right|
			+ \mathcal{P}_n
\\
	{p}_n(\mat{x}) &= \sqrt{2m|V(\mat{x})-E|}
\\
	\kappa_i &= \frac{V_n(\mat{x}_i) - V_n(\mat{x}_{i-1})}{|\mat{x}_i - \mat{x}_{i-1}|},
\end{align}
where between each bead
we have used the analytical expression for the abbreviated action in a linear potential,
and the factor of two accounts for the return journey of the trajectory.
The absolute value of the momentum is taken such that the function
returns real values even when beads stray into the classically allowed region.
This ensures that the function is smooth and well-defined everywhere as is required by most optimization routines.
The final optimized pathway should however lie entirely in the classically forbidden region.
This requirement can be easily checked.

In this formulation,
it is necessary to 
ensure that the beads remain evenly spaced
without biasing the instanton pathway.
The simplest way to achieve this is to
include a penalty function,
\begin{align}
	\mathcal{P}_n &= \chi_n \sum_{i=1}^{N_n/2} \big(\delta x_i  - \braket{\delta x} \! \big)^2
\\
	\delta x_i &= |\mat{x}_i-\mat{x}_{i-1}|
\\
	\braket{\delta x} &= \frac{2}{N_n} \sum_{i=1}^{N_n/2} \delta x_i.
\end{align}
This type of approach has been applied successfully
to locate folding pathways in proteins. \cite{Faccioli2006pathways,*Beccara2012folding}
However, alternative methods based on generalizations of the nudged-elastic-band algorithm
avoid using penalty functions and may be more efficient. \cite{Einarsdottir2012path}
The value of the scalar $\chi_n$ should not affect the result of a converged optimization
and can be chosen by the user to maximize efficiency.

As in all optimization problems, a good initial guess is required to ensure fast convergence to the global minimum.
Instanton optimizations are best performed in stages with increasing numbers of beads and decreasing temperatures. \cite{tunnel}
An equally spaced straight line normal to the crossing seam
provides a reasonable starting point at high temperatures.

The imaginary time intervals between each bead are evaluated
from the derivative of the abbreviated action with respect to energy
as
\begin{align}
	\delta \tau_i = \left|\frac{{p}_n(\mat{x}_{i-1}) - {p}_n(\mat{x}_i)}{\kappa_i}\right|.
	\label{dt}
\end{align}
Thus the total imaginary time along each trajectory is $\tau_n=2\sum_{i=1}^{N_n/2} \delta\tau_i$
and we define \mbox{$\tau=\beta\hbar/(1+\tau_0/\tau_1)$}.

Classical trajectories could be located by optimizing the abbreviated action \eqn{Wn} for a given energy.
This approach would give the microcanonical instanton rates discussed in Paper I\@.
However, it is the thermal rate which is of most interest, for which the value of $E$ is not known \textit{a priori}.
We therefore use the value of the full action in the Hamilton-Jacobi picture,
\begin{multline}
	{S}(\mathbf{x},E) = {W}_0(\mat{x}_{N_0/2},\dots,\mat{x}_{N_0};E) \\+ {W}_1(\mat{x}_{N_0},\dots,\mat{x}_{N_0+N_1/2};E) + \beta E.
\end{multline}
This function is minimized with respect to
the independent beads $\mathbf{x}=\{\mat{x}_{N_0/2},\dots,\mat{x}_{N_0+N_1/2}\}$
and energy simultaneously
under the constraint that the pathways terminate at a turning point,
i.e.\
$V_0(\mat{x}_{N_0/2}) = E$ and $V_1(\mat{x}_{N_0+N_1/2}) = E$.
Constrained optimization methods such as sequential least squares programming
are ideal for this task.

This Hamilton-Jacobi approach to locating instantons
has significant advantages over the standard ring-polymer instanton approach,
where the beads tend to accumulate near turning points. \cite{Rommel2011grids}
By forcing the beads to be evenly spaced along each trajectory,
we expect that fewer beads will be required to converge the action integral.
The convergence is further improved by using the scheme based on the analytic result for linear potentials.
Another advantage is that the standard instanton-finding methods
employ a saddle-point search, \cite{rpinst}
whereas the new approach requires only a minimization.
It is usually far less computationally demanding to locate the latter type of stationary point.

However, it is known that the evenly spaced pathway does not give good estimates for the instanton prefactor, \cite{Rommel2011grids}
even when $N$ is large enough to converge the action to a high accuracy.
This was also confirmed by our own numerical tests, employing the formulae in \secref{derivatives} with \eqn{dt}.
It seems that the ring-polymer instanton methods described in Secs.~\ref{sec:RPI} and \ref{sec:event} with equal time-steps is better for computing the derivatives
whereas this Hamilton-Jacobi method with evenly spaced beads is better for estimating the action.

We therefore propose that the following combination of the methods presented above is used for computing the rate:
\begin{itemize}
\item
The Hamilton-Jacobi method can be used to locate the instanton pathway and find the stationary value of $\tau$.
We also take the action, $\tilde{S}$, from this calculation.
\item
Using cubic spline interpolation along the imaginary time coordinate, \cite{NumRep}
the two trajectories are modified to give equal time-steps along each trajectory,
and the resulting pathway minimized, keeping $\tau$ fixed.
\item
The remaining beads in the two bounce trajectories are obtained by symmetry
and the derivatives of the actions, $\tilde{S}_0$ and $\tilde{S}_1$, computed using the formulae in \secref{derivatives}.
\item
The rate constant can be then be evaluated using \eqn{kSpartial}.
\end{itemize}

In \secref{results}, we apply this combined method and compare it with the saddle-point search of \secref{event} to a model problem.

\section{Application to a model system}
\label{sec:results}

We consider a numerical application of the golden-rule instanton method
to a spin-boson model of electron transfer.  \cite{Garg1985spinboson,*Leggett1987spinboson,Weiss}
Note that the methods are also directly applicable to anharmonic systems,
but here we intend to compare with the exact results, which are easily available only for integrable systems.

The spin-boson model was defined in Paper I and we use the same notation here
with parameters chosen to describe condensed-phase electron transfer at typical conditions.
The temperature is
$T=300\,\mathrm{K}$,
and the spectral density of the bath has Debye form
$J(\omega) = \frac{\lambda}{2}\frac{\omega\omega_\mathrm{c}}{\omega^2+\omega_\mathrm{c}^2}$,
with the characteristic frequency
$\omega_\mathrm{c}=500\,\mathrm{cm}^{-1}$,
and reorganization energy
$\lambda=40\,\mathrm{kcal/mol}$.
The spectral density is discretized with $f=12$ bath modes using 
\cite{Wang2003RuRu,Berkelbach2012hybrid}
\begin{align}
	\omega_j &= \omega_\mathrm{c} \tan\frac{\left(j-\half\right)\pi}{2f}
\\
	c_j &= \sqrt\frac{\lambda}{2f} \, \omega_j,
\end{align}
where $j\in\{1,\dots,f\}$.
We include a bias to products of 
$\epsilon=10\,\mathrm{kcal/mol}$.

The electronic coupling, $\Delta$, is constant,
but for the purposes of generality we do not specify its value.
It must of course be small enough that the golden-rule approximation is valid.
Results are presented relative to the classical rate such that they are dimensionless and do not depend on $\Delta$.
It was found that 12 bath modes are enough to converge the ratio to less than 2\%.

For this model, the classical rate is given by Marcus theory as \cite{Marcus1985review}
\begin{align}
	k_\text{cl} = \frac{\Delta^2}{\hbar} \sqrt\frac{\pi\beta}{\lambda} \, \eu{-\beta(\lambda-\epsilon)^2/4\lambda}.
\end{align}
Formulae presented in Paper I give
the semiclassical golden-rule rate, $k_\mathrm{SC}$,
with $\tau$ obtained numerically by a one-dimensional maximization,
as 36.3 $k_\mathrm{cl}$.
This is close to the quantum golden-rule rate,
which was found to be 36.6 $k_\mathrm{cl}$ by numerical integration.
Here, as was also observed in \Ref{Bader1990golden}, nuclear tunnelling has a significant effect on the rate.

The two numerical approaches outlined in \secref{numerical} were applied to the model for various numbers of ring-polymer beads.
In each case, the starting point for new instanton searches
was given by a spline interpolation \cite{NumRep}
of the trajectories from previous optimizations with fewer beads.
The results are given in Table \ref{table}.

\begin{table}
\caption{Results for the two numerical methods, Lagrangian and Hamilton-Jacobi (Ham-Jac), described in Secs.~\ref{sec:event} and \ref{sec:HamJac}.
In both cases we take $N_1/N_0\simeq0.3$, although ensuring that $N_0$ and $N_1$ are even.
The semiclassical instanton results are given in the final row, computed using formulae from Paper I.}
\label{table}
\begin{ruledtabular}
\begin{tabular}{d{3.0}d{1.3}d{0.4}d{2.1}d{1.3}d{2.1}}
	    & \multicolumn{3}{c}{Lagrangian} & \multicolumn{2}{c}{Ham-Jac} \\
\cline{2-4} \cline{5-6}
	  \head{$N$} & \head{$\tilde{S}/\hbar$} & \head{$\tau/\beta\hbar$} & \head{$k_\text{SC}/k_\text{cl}$} & \head{$\tilde{S}/\hbar$} & \head{$k_\text{SC}/k_\text{cl}$} \\
\hline
	  8 & 6.558 & 0.3248 & 23.2 & 6.152 & 31.1 \\
	 16 & 6.179 & 0.3163 & 33.4 & 6.051 & 36.5 \\
	 32 & 6.058 & 0.3131 & 36.3 & 6.020 & 36.0 \\
	 64 & 6.022 & 0.3119 & 36.1 & 6.013 & 36.2 \\
	128 & 6.013 & 0.3117 & 36.2 & 6.012 & 36.3 \\
	256 & 6.012 & 0.3116 & 36.3 & 6.011 & 36.3 \\
\hline
	\infty & 6.011 & 0.3116 & 36.3 & 6.011 & 36.3 \\
\end{tabular}
\end{ruledtabular}
\end{table}

As expected, the rates obtained by both numerical methods tend to the semiclassical results in the large $N$ limit.
The Hamilton-Jacobi formulation is found to give better estimates of $\tilde{S}$ than the Lagrangian formulation for the same number of beads.
Using the combined approach in which this action is used alongside the derivatives found from an optimized instanton with equal time-steps,
requires in each case about half as many beads for the same error in the rate.
This would lead to a significant advantage when treating more complex systems.

\section{Conclusions}
\label{sec:conclusions}

In this paper we have described a ring-polymer formulation
of the golden-rule instanton approach derived in Paper I\@. \cite{GoldenGreens}
This formulation is amenable to efficient numerical evaluation
and we have suggested two methods for its computation.

The method based on the Hamilton-Jacobi formalism
appears to be more efficient at obtaining the instanton trajectory and its action.
This approach forces the energy along both instanton trajectories to be equal,
which is a fundamental aspect of our time-independent derivation.
Similar approaches may also prove efficient for locating instantons used in other calculations,
such as adiabatic rate theory and tunnelling splitting calculations.

The ring-polymer instanton was shown to be equivalent to
a steepest-descent evaluation of Wolynes' quantum instanton approach, \cite{Wolynes1987nonadiabatic}
thus providing a strong link between the two methods.
Quantum instanton approaches \cite{Vanicek2005QI}
employ a Gaussian approximation to the flux-flux correlation function
whose short-time behaviour is computed using exact path-integral methods.
Notable deviations from Gaussian behaviour occur even for the simplest problem of free-particle propagation \cite{Miller2003QI}
and
it seems that the flux-flux correlation function cannot be assumed to be Gaussian
if a rate theory is required which gives a good approximation to the high-temperature limit.
The golden-rule instanton method does not however suffer from these problems.

All instanton methods will fail when the potential-energy surfaces exhibit oscillations,
as occurs with liquids,
such that many minima appear on the ring-polymer surface.
In this case, the steepest-descent integrals employed in the instanton derivation are not valid
and path-integral sampling methods such as Wolynes' approach, \eqn{Wolynes}, are necessary.
However, for systems where the environment is not fluxional,
such as in solids \cite{EsquinaziBook} or certain gas-phase molecules,
the instanton approach may be more accurate
as well as much more efficient.

\section{Acknowledgement}
The author gratefully acknowledges a Research Fellowship from the Alexander von Humboldt Foundation
and would like to thank Michael Thoss for helpful comments on the manuscript.

\bibliography{references} %

\begin{thebibliography}{72}%
\makeatletter
\providecommand \@ifxundefined [1]{%
 \@ifx{#1\undefined}
}%
\providecommand \@ifnum [1]{%
 \ifnum #1\expandafter \@firstoftwo
 \else \expandafter \@secondoftwo
 \fi
}%
\providecommand \@ifx [1]{%
 \ifx #1\expandafter \@firstoftwo
 \else \expandafter \@secondoftwo
 \fi
}%
\providecommand \natexlab [1]{#1}%
\providecommand \enquote  [1]{``#1''}%
\providecommand \bibnamefont  [1]{#1}%
\providecommand \bibfnamefont [1]{#1}%
\providecommand \citenamefont [1]{#1}%
\providecommand \href@noop [0]{\@secondoftwo}%
\providecommand \href [0]{\begingroup \@sanitize@url \@href}%
\providecommand \@href[1]{\@@startlink{#1}\@@href}%
\providecommand \@@href[1]{\endgroup#1\@@endlink}%
\providecommand \@sanitize@url [0]{\catcode `\\12\catcode `\$12\catcode
  `\&12\catcode `\#12\catcode `\^12\catcode `\_12\catcode `\%12\relax}%
\providecommand \@@startlink[1]{}%
\providecommand \@@endlink[0]{}%
\providecommand \url  [0]{\begingroup\@sanitize@url \@url }%
\providecommand \@url [1]{\endgroup\@href {#1}{\urlprefix }}%
\providecommand \urlprefix  [0]{URL }%
\providecommand \Eprint [0]{\href }%
\providecommand \doibase [0]{http://dx.doi.org/}%
\providecommand \selectlanguage [0]{\@gobble}%
\providecommand \bibinfo  [0]{\@secondoftwo}%
\providecommand \bibfield  [0]{\@secondoftwo}%
\providecommand \translation [1]{[#1]}%
\providecommand \BibitemOpen [0]{}%
\providecommand \bibitemStop [0]{}%
\providecommand \bibitemNoStop [0]{.\EOS\space}%
\providecommand \EOS [0]{\spacefactor3000\relax}%
\providecommand \BibitemShut  [1]{\csname bibitem#1\endcsname}%
\let\auto@bib@innerbib\@empty
\bibitem [{\citenamefont {Richardson}, \citenamefont {Bauer},\ and\
  \citenamefont {Thoss}()}]{GoldenGreens}%
  \BibitemOpen
  \bibfield  {author} {\bibinfo {author} {\bibfnamefont {J.~O.}\ \bibnamefont
  {Richardson}}, \bibinfo {author} {\bibfnamefont {R.}~\bibnamefont {Bauer}}, \
  and\ \bibinfo {author} {\bibfnamefont {M.}~\bibnamefont {Thoss}},\
  }\href@noop {} {\ }\Eprint {http://arxiv.org/abs/1508.04919}
  {arXiv:1508.04919 [physics.chem-ph]} \BibitemShut {NoStop}%
\bibitem [{\citenamefont {Richardson}\ and\ \citenamefont
  {Althorpe}(2009)}]{rpinst}%
  \BibitemOpen
  \bibfield  {author} {\bibinfo {author} {\bibfnamefont {J.~O.}\ \bibnamefont
  {Richardson}}\ and\ \bibinfo {author} {\bibfnamefont {S.~C.}\ \bibnamefont
  {Althorpe}},\ }\href {\doibase 10.1063/1.3267318} {\bibfield  {journal}
  {\bibinfo  {journal} {J.~Chem. Phys.}\ }\textbf {\bibinfo {volume} {131}},\
  \bibinfo {pages} {214106} (\bibinfo {year} {2009})}\BibitemShut {NoStop}%
\bibitem [{\citenamefont {Andersson}\ \emph {et~al.}(2009)\citenamefont
  {Andersson}, \citenamefont {Nyman}, \citenamefont {Arnaldsson}, \citenamefont
  {Manthe},\ and\ \citenamefont {J{\'o}nsson}}]{Andersson2009Hmethane}%
  \BibitemOpen
  \bibfield  {author} {\bibinfo {author} {\bibfnamefont {S.}~\bibnamefont
  {Andersson}}, \bibinfo {author} {\bibfnamefont {G.}~\bibnamefont {Nyman}},
  \bibinfo {author} {\bibfnamefont {A.}~\bibnamefont {Arnaldsson}}, \bibinfo
  {author} {\bibfnamefont {U.}~\bibnamefont {Manthe}}, \ and\ \bibinfo {author}
  {\bibfnamefont {H.}~\bibnamefont {J{\'o}nsson}},\ }\href {\doibase
  10.1021/jp811070w} {\bibfield  {journal} {\bibinfo  {journal} {J.~Phys.
  Chem.~A}\ }\textbf {\bibinfo {volume} {113}},\ \bibinfo {pages} {4468}
  (\bibinfo {year} {2009})}\BibitemShut {NoStop}%
\bibitem [{\citenamefont {Andersson}, \citenamefont {Goumans},\ and\
  \citenamefont {Arnaldsson}(2011)}]{Andersson2011HCO}%
  \BibitemOpen
  \bibfield  {author} {\bibinfo {author} {\bibfnamefont {S.}~\bibnamefont
  {Andersson}}, \bibinfo {author} {\bibfnamefont {T.~P.~M.}\ \bibnamefont
  {Goumans}}, \ and\ \bibinfo {author} {\bibfnamefont {A.}~\bibnamefont
  {Arnaldsson}},\ }\href {\doibase 10.1016/j.cplett.2011.07.073} {\bibfield
  {journal} {\bibinfo  {journal} {Chem. Phys. Lett.}\ }\textbf {\bibinfo
  {volume} {513}},\ \bibinfo {pages} {31} (\bibinfo {year} {2011})}\BibitemShut
  {NoStop}%
\bibitem [{\citenamefont {Goumans}\ and\ \citenamefont
  {K{\"a}stner}(2010)}]{Goumans2010Hbenzene}%
  \BibitemOpen
  \bibfield  {author} {\bibinfo {author} {\bibfnamefont {T.~P.~M.}\
  \bibnamefont {Goumans}}\ and\ \bibinfo {author} {\bibfnamefont
  {J.}~\bibnamefont {K{\"a}stner}},\ }\href {\doibase 10.1002/anie.201001311}
  {\bibfield  {journal} {\bibinfo  {journal} {Angew. Chem. Int. Edit.}\
  }\textbf {\bibinfo {volume} {49}},\ \bibinfo {pages} {7350} (\bibinfo {year}
  {2010})}\BibitemShut {NoStop}%
\bibitem [{\citenamefont {Goumans}\ and\ \citenamefont
  {K{\"a}stner}(2011)}]{Goumans2011Hmethanol}%
  \BibitemOpen
  \bibfield  {author} {\bibinfo {author} {\bibfnamefont {T.~P.~M.}\
  \bibnamefont {Goumans}}\ and\ \bibinfo {author} {\bibfnamefont
  {J.}~\bibnamefont {K{\"a}stner}},\ }\href {\doibase 10.1021/jp206048f}
  {\bibfield  {journal} {\bibinfo  {journal} {J.~Phys. Chem.~A}\ }\textbf
  {\bibinfo {volume} {115}},\ \bibinfo {pages} {10767} (\bibinfo {year}
  {2011})}\BibitemShut {NoStop}%
\bibitem [{\citenamefont {Meisner}, \citenamefont {Rommel},\ and\ \citenamefont
  {K{\"a}stner}(2011)}]{Meisner2011isotope}%
  \BibitemOpen
  \bibfield  {author} {\bibinfo {author} {\bibfnamefont {J.}~\bibnamefont
  {Meisner}}, \bibinfo {author} {\bibfnamefont {J.~B.}\ \bibnamefont {Rommel}},
  \ and\ \bibinfo {author} {\bibfnamefont {J.}~\bibnamefont {K{\"a}stner}},\
  }\href {\doibase 10.1002/jcc.21930} {\bibfield  {journal} {\bibinfo
  {journal} {J.~Comput. Chem.}\ }\textbf {\bibinfo {volume} {32}},\ \bibinfo
  {pages} {3456} (\bibinfo {year} {2011})}\BibitemShut {NoStop}%
\bibitem [{\citenamefont {Rommel}\ \emph {et~al.}(2012)\citenamefont {Rommel},
  \citenamefont {Liu}, \citenamefont {Werner},\ and\ \citenamefont
  {K\"astner}}]{Rommel2012enzyme}%
  \BibitemOpen
  \bibfield  {author} {\bibinfo {author} {\bibfnamefont {J.~B.}\ \bibnamefont
  {Rommel}}, \bibinfo {author} {\bibfnamefont {Y.}~\bibnamefont {Liu}},
  \bibinfo {author} {\bibfnamefont {H.-J.}\ \bibnamefont {Werner}}, \ and\
  \bibinfo {author} {\bibfnamefont {J.}~\bibnamefont {K\"astner}},\ }\href
  {\doibase 10.1021/jp308526t} {\bibfield  {journal} {\bibinfo  {journal}
  {J.~Phys.\ Chem.~B}\ }\textbf {\bibinfo {volume} {116}},\ \bibinfo {pages}
  {13682} (\bibinfo {year} {2012})}\BibitemShut {NoStop}%
\bibitem [{\citenamefont {Rommel}, \citenamefont {Goumans},\ and\ \citenamefont
  {K\"astner}(2011)}]{Rommel2011locating}%
  \BibitemOpen
  \bibfield  {author} {\bibinfo {author} {\bibfnamefont {J.~B.}\ \bibnamefont
  {Rommel}}, \bibinfo {author} {\bibfnamefont {T.~P.~M.}\ \bibnamefont
  {Goumans}}, \ and\ \bibinfo {author} {\bibfnamefont {J.}~\bibnamefont
  {K\"astner}},\ }\href {\doibase 10.1021/ct100658y} {\bibfield  {journal}
  {\bibinfo  {journal} {J.~Chem. Theory Comput.}\ }\textbf {\bibinfo {volume}
  {7}},\ \bibinfo {pages} {690} (\bibinfo {year} {2011})}\BibitemShut {NoStop}%
\bibitem [{\citenamefont {Rommel}\ and\ \citenamefont
  {K{\"a}stner}(2011)}]{Rommel2011grids}%
  \BibitemOpen
  \bibfield  {author} {\bibinfo {author} {\bibfnamefont {J.~B.}\ \bibnamefont
  {Rommel}}\ and\ \bibinfo {author} {\bibfnamefont {J.}~\bibnamefont
  {K{\"a}stner}},\ }\href {\doibase 10.1063/1.3587240} {\bibfield  {journal}
  {\bibinfo  {journal} {J.~Chem. Phys.}\ }\textbf {\bibinfo {volume} {134}},\
  \bibinfo {pages} {184107} (\bibinfo {year} {2011})}\BibitemShut {NoStop}%
\bibitem [{\citenamefont {Althorpe}(2011)}]{Althorpe2011ImF}%
  \BibitemOpen
  \bibfield  {author} {\bibinfo {author} {\bibfnamefont {S.~C.}\ \bibnamefont
  {Althorpe}},\ }\href {\doibase 10.1063/1.3563045} {\bibfield  {journal}
  {\bibinfo  {journal} {J.~Chem. Phys.}\ }\textbf {\bibinfo {volume} {134}},\
  \bibinfo {pages} {114104} (\bibinfo {year} {2011})}\BibitemShut {NoStop}%
\bibitem [{\citenamefont {P{\'e}rez~de Tudela}\ \emph
  {et~al.}(2014)\citenamefont {P{\'e}rez~de Tudela}, \citenamefont
  {Suleimanov}, \citenamefont {Richardson}, \citenamefont
  {S{\'a}ez~R{\'a}banos}, \citenamefont {Green},\ and\ \citenamefont
  {Aoiz}}]{DMuH}%
  \BibitemOpen
  \bibfield  {author} {\bibinfo {author} {\bibfnamefont {R.}~\bibnamefont
  {P{\'e}rez~de Tudela}}, \bibinfo {author} {\bibfnamefont {Y.~V.}\
  \bibnamefont {Suleimanov}}, \bibinfo {author} {\bibfnamefont {J.~O.}\
  \bibnamefont {Richardson}}, \bibinfo {author} {\bibfnamefont
  {V.}~\bibnamefont {S{\'a}ez~R{\'a}banos}}, \bibinfo {author} {\bibfnamefont
  {W.~H.}\ \bibnamefont {Green}}, \ and\ \bibinfo {author} {\bibfnamefont
  {F.~J.}\ \bibnamefont {Aoiz}},\ }\href {\doibase 10.1021/jz502216g}
  {\bibfield  {journal} {\bibinfo  {journal} {J. Phys. Chem. Lett.}\ }\textbf
  {\bibinfo {volume} {5}},\ \bibinfo {pages} {4219} (\bibinfo {year}
  {2014})}\BibitemShut {NoStop}%
\bibitem [{\citenamefont {Richardson}\ and\ \citenamefont
  {Althorpe}(2011)}]{tunnel}%
  \BibitemOpen
  \bibfield  {author} {\bibinfo {author} {\bibfnamefont {J.~O.}\ \bibnamefont
  {Richardson}}\ and\ \bibinfo {author} {\bibfnamefont {S.~C.}\ \bibnamefont
  {Althorpe}},\ }\href {\doibase 10.1063/1.3530589} {\bibfield  {journal}
  {\bibinfo  {journal} {J.~Chem. Phys.}\ }\textbf {\bibinfo {volume} {134}},\
  \bibinfo {pages} {054109} (\bibinfo {year} {2011})}\BibitemShut {NoStop}%
\bibitem [{\citenamefont {Richardson}, \citenamefont {Althorpe},\ and\
  \citenamefont {Wales}(2011)}]{water}%
  \BibitemOpen
  \bibfield  {author} {\bibinfo {author} {\bibfnamefont {J.~O.}\ \bibnamefont
  {Richardson}}, \bibinfo {author} {\bibfnamefont {S.~C.}\ \bibnamefont
  {Althorpe}}, \ and\ \bibinfo {author} {\bibfnamefont {D.~J.}\ \bibnamefont
  {Wales}},\ }\href {\doibase 10.1063/1.3640429} {\bibfield  {journal}
  {\bibinfo  {journal} {J.~Chem. Phys.}\ }\textbf {\bibinfo {volume} {135}},\
  \bibinfo {pages} {124109} (\bibinfo {year} {2011})}\BibitemShut {NoStop}%
\bibitem [{\citenamefont {Richardson}\ \emph {et~al.}(2013)\citenamefont
  {Richardson}, \citenamefont {Wales}, \citenamefont {Althorpe}, \citenamefont
  {McLaughlin}, \citenamefont {Viant}, \citenamefont {Shih},\ and\
  \citenamefont {Saykally}}]{octamer}%
  \BibitemOpen
  \bibfield  {author} {\bibinfo {author} {\bibfnamefont {J.~O.}\ \bibnamefont
  {Richardson}}, \bibinfo {author} {\bibfnamefont {D.~J.}\ \bibnamefont
  {Wales}}, \bibinfo {author} {\bibfnamefont {S.~C.}\ \bibnamefont {Althorpe}},
  \bibinfo {author} {\bibfnamefont {R.~P.}\ \bibnamefont {McLaughlin}},
  \bibinfo {author} {\bibfnamefont {M.~R.}\ \bibnamefont {Viant}}, \bibinfo
  {author} {\bibfnamefont {O.}~\bibnamefont {Shih}}, \ and\ \bibinfo {author}
  {\bibfnamefont {R.~J.}\ \bibnamefont {Saykally}},\ }\href {\doibase
  10.1021/jp311306a} {\bibfield  {journal} {\bibinfo  {journal} {J.~Phys.
  Chem.~A}\ }\textbf {\bibinfo {volume} {117}},\ \bibinfo {pages} {6960 }
  (\bibinfo {year} {2013})}\BibitemShut {NoStop}%
\bibitem [{\citenamefont {Kawatsu}\ and\ \citenamefont
  {Miura}(2014)}]{Kawatsu2014RPI}%
  \BibitemOpen
  \bibfield  {author} {\bibinfo {author} {\bibfnamefont {T.}~\bibnamefont
  {Kawatsu}}\ and\ \bibinfo {author} {\bibfnamefont {S.}~\bibnamefont
  {Miura}},\ }\href {\doibase 10.1063/1.4885437} {\bibfield  {journal}
  {\bibinfo  {journal} {J.~Chem. Phys.}\ }\textbf {\bibinfo {volume} {141}},\
  \bibinfo {pages} {024101} (\bibinfo {year} {2014})}\BibitemShut {NoStop}%
\bibitem [{\citenamefont {Kawatsu}\ and\ \citenamefont
  {Miura}(2015)}]{Kawatsu2015NH3}%
  \BibitemOpen
  \bibfield  {author} {\bibinfo {author} {\bibfnamefont {T.}~\bibnamefont
  {Kawatsu}}\ and\ \bibinfo {author} {\bibfnamefont {S.}~\bibnamefont
  {Miura}},\ }\href {\doibase 10.1016/j.cplett.2015.05.072} {\bibfield
  {journal} {\bibinfo  {journal} {Chem. Phys. Lett.}\ }\textbf {\bibinfo
  {volume} {634}},\ \bibinfo {pages} {146} (\bibinfo {year}
  {2015})}\BibitemShut {NoStop}%
\bibitem [{\citenamefont {Chapman}, \citenamefont {Garrett},\ and\
  \citenamefont {Miller}(1975)}]{Chapman1975rates}%
  \BibitemOpen
  \bibfield  {author} {\bibinfo {author} {\bibfnamefont {S.}~\bibnamefont
  {Chapman}}, \bibinfo {author} {\bibfnamefont {B.~C.}\ \bibnamefont
  {Garrett}}, \ and\ \bibinfo {author} {\bibfnamefont {W.~H.}\ \bibnamefont
  {Miller}},\ }\href {\doibase 10.1063/1.431620} {\bibfield  {journal}
  {\bibinfo  {journal} {J.~Chem. Phys.}\ }\textbf {\bibinfo {volume} {63}},\
  \bibinfo {pages} {2710} (\bibinfo {year} {1975})}\BibitemShut {NoStop}%
\bibitem [{\citenamefont {Benderskii}, \citenamefont {Makarov},\ and\
  \citenamefont {Wight}(1994)}]{Benderskii}%
  \BibitemOpen
  \bibfield  {author} {\bibinfo {author} {\bibfnamefont {V.~A.}\ \bibnamefont
  {Benderskii}}, \bibinfo {author} {\bibfnamefont {D.~E.}\ \bibnamefont
  {Makarov}}, \ and\ \bibinfo {author} {\bibfnamefont {C.~A.}\ \bibnamefont
  {Wight}},\ }\href@noop {} {\emph {\bibinfo {title} {Chemical Dynamics at Low
  Temperatures}}},\ \bibinfo {series} {Adv. Chem. Phys.}, Vol.~\bibinfo
  {volume} {88}\ (\bibinfo  {publisher} {Wiley},\ \bibinfo {address} {New
  York},\ \bibinfo {year} {1994})\BibitemShut {NoStop}%
\bibitem [{\citenamefont {Wang}, \citenamefont {Thoss},\ and\ \citenamefont
  {Miller}(2001)}]{Wang2001hybrid}%
  \BibitemOpen
  \bibfield  {author} {\bibinfo {author} {\bibfnamefont {H.}~\bibnamefont
  {Wang}}, \bibinfo {author} {\bibfnamefont {M.}~\bibnamefont {Thoss}}, \ and\
  \bibinfo {author} {\bibfnamefont {W.~H.}\ \bibnamefont {Miller}},\ }\href
  {\doibase 10.1063/1.1385561} {\bibfield  {journal} {\bibinfo  {journal}
  {J.~Chem. Phys.}\ }\textbf {\bibinfo {volume} {115}},\ \bibinfo {pages}
  {2979} (\bibinfo {year} {2001})}\BibitemShut {NoStop}%
\bibitem [{\citenamefont {Thoss}, \citenamefont {Wang},\ and\ \citenamefont
  {Miller}(2001)}]{Thoss2001hybrid}%
  \BibitemOpen
  \bibfield  {author} {\bibinfo {author} {\bibfnamefont {M.}~\bibnamefont
  {Thoss}}, \bibinfo {author} {\bibfnamefont {H.}~\bibnamefont {Wang}}, \ and\
  \bibinfo {author} {\bibfnamefont {W.~H.}\ \bibnamefont {Miller}},\ }\href
  {\doibase 10.1063/1.1385562} {\bibfield  {journal} {\bibinfo  {journal}
  {J.~Chem. Phys.}\ }\textbf {\bibinfo {volume} {115}},\ \bibinfo {pages}
  {2991} (\bibinfo {year} {2001})}\BibitemShut {NoStop}%
\bibitem [{\citenamefont {Wang}, \citenamefont {Skinner},\ and\ \citenamefont
  {Thoss}(2006)}]{Wang2006flux}%
  \BibitemOpen
  \bibfield  {author} {\bibinfo {author} {\bibfnamefont {H.}~\bibnamefont
  {Wang}}, \bibinfo {author} {\bibfnamefont {D.~E.}\ \bibnamefont {Skinner}}, \
  and\ \bibinfo {author} {\bibfnamefont {M.}~\bibnamefont {Thoss}},\ }\href
  {\doibase 10.1063/1.2363195} {\bibfield  {journal} {\bibinfo  {journal}
  {J.~Chem. Phys.}\ }\textbf {\bibinfo {volume} {125}},\ \bibinfo {pages}
  {174502} (\bibinfo {year} {2006})}\BibitemShut {NoStop}%
\bibitem [{\citenamefont {Marchi}\ and\ \citenamefont
  {Chandler}(1991)}]{Marchi1991tunnelling}%
  \BibitemOpen
  \bibfield  {author} {\bibinfo {author} {\bibfnamefont {M.}~\bibnamefont
  {Marchi}}\ and\ \bibinfo {author} {\bibfnamefont {D.}~\bibnamefont
  {Chandler}},\ }\href {\doibase 10.1063/1.461096} {\bibfield  {journal}
  {\bibinfo  {journal} {J.~Chem. Phys.}\ }\textbf {\bibinfo {volume} {95}},\
  \bibinfo {pages} {889} (\bibinfo {year} {1991})}\BibitemShut {NoStop}%
\bibitem [{\citenamefont {Topaler}\ and\ \citenamefont
  {Makri}(1996)}]{Topaler1996nonadiabatic}%
  \BibitemOpen
  \bibfield  {author} {\bibinfo {author} {\bibfnamefont {M.}~\bibnamefont
  {Topaler}}\ and\ \bibinfo {author} {\bibfnamefont {N.}~\bibnamefont
  {Makri}},\ }\href {\doibase 10.1021/jp951673k} {\bibfield  {journal}
  {\bibinfo  {journal} {J.~Phys. Chem.}\ }\textbf {\bibinfo {volume} {100}},\
  \bibinfo {pages} {4430} (\bibinfo {year} {1996})}\BibitemShut {NoStop}%
\bibitem [{\citenamefont {M{\"u}hlbacher}\ and\ \citenamefont
  {Egger}(2003)}]{Muehlbacher2003spinboson}%
  \BibitemOpen
  \bibfield  {author} {\bibinfo {author} {\bibfnamefont {L.}~\bibnamefont
  {M{\"u}hlbacher}}\ and\ \bibinfo {author} {\bibfnamefont {R.}~\bibnamefont
  {Egger}},\ }\href {\doibase 10.1063/1.1523014} {\bibfield  {journal}
  {\bibinfo  {journal} {J.~Chem. Phys.}\ }\textbf {\bibinfo {volume} {118}},\
  \bibinfo {pages} {179} (\bibinfo {year} {2003})}\BibitemShut {NoStop}%
\bibitem [{\citenamefont {M{\"u}hlbacher}\ and\ \citenamefont
  {Egger}(2004)}]{Muehlbacher2004asymmetric}%
  \BibitemOpen
  \bibfield  {author} {\bibinfo {author} {\bibfnamefont {L.}~\bibnamefont
  {M{\"u}hlbacher}}\ and\ \bibinfo {author} {\bibfnamefont {R.}~\bibnamefont
  {Egger}},\ }\href {\doibase 10.1016/j.chemphys.2003.08.021} {\bibfield
  {journal} {\bibinfo  {journal} {Chem. Phys.}\ }\textbf {\bibinfo {volume}
  {296}},\ \bibinfo {pages} {193} (\bibinfo {year} {2004})}\BibitemShut
  {NoStop}%
\bibitem [{\citenamefont {Ananth}\ and\ \citenamefont
  {Miller~III}(2012)}]{Ananth2012QUAPI}%
  \BibitemOpen
  \bibfield  {author} {\bibinfo {author} {\bibfnamefont {N.}~\bibnamefont
  {Ananth}}\ and\ \bibinfo {author} {\bibfnamefont {T.~F.}\ \bibnamefont
  {Miller~III}},\ }\href {\doibase 10.1080/00268976.2012.686641} {\bibfield
  {journal} {\bibinfo  {journal} {Mol. Phys.}\ }\textbf {\bibinfo {volume}
  {110}},\ \bibinfo {pages} {1009} (\bibinfo {year} {2012})}\BibitemShut
  {NoStop}%
\bibitem [{\citenamefont {Huo}, \citenamefont {Miller~III},\ and\ \citenamefont
  {Coker}(2013)}]{Huo2013PLDM}%
  \BibitemOpen
  \bibfield  {author} {\bibinfo {author} {\bibfnamefont {P.}~\bibnamefont
  {Huo}}, \bibinfo {author} {\bibfnamefont {T.~F.}\ \bibnamefont {Miller~III}},
  \ and\ \bibinfo {author} {\bibfnamefont {D.~F.}\ \bibnamefont {Coker}},\
  }\href {\doibase 10.1063/1.4826163} {\bibfield  {journal} {\bibinfo
  {journal} {J.~Chem. Phys.}\ }\textbf {\bibinfo {volume} {139}},\ \bibinfo
  {pages} {151103} (\bibinfo {year} {2013})}\BibitemShut {NoStop}%
\bibitem [{\citenamefont {Huo}\ and\ \citenamefont
  {Miller~III}(2015)}]{Huo2015PLDM}%
  \BibitemOpen
  \bibfield  {author} {\bibinfo {author} {\bibfnamefont {P.}~\bibnamefont
  {Huo}}\ and\ \bibinfo {author} {\bibfnamefont {T.~F.}\ \bibnamefont
  {Miller~III}},\ }\href {\doibase 10.1039/c5cp02517f} {\bibfield  {journal}
  {\bibinfo  {journal} {Phys. Chem. Chem. Phys.}\ } (\bibinfo {year} {2015}),\
  10.1039/c5cp02517f}\BibitemShut {NoStop}%
\bibitem [{\citenamefont {Kapral}(2015)}]{Kapral2015QCL}%
  \BibitemOpen
  \bibfield  {author} {\bibinfo {author} {\bibfnamefont {R.}~\bibnamefont
  {Kapral}},\ }\href {\doibase 10.1088/0953-8984/27/7/073201} {\bibfield
  {journal} {\bibinfo  {journal} {J. Phys.-Condens. Mat.}\ }\textbf {\bibinfo
  {volume} {27}},\ \bibinfo {pages} {073201} (\bibinfo {year}
  {2015})}\BibitemShut {NoStop}%
\bibitem [{\citenamefont {Sun}, \citenamefont {Wang},\ and\ \citenamefont
  {Miller}(1998)}]{Sun1998mapping}%
  \BibitemOpen
  \bibfield  {author} {\bibinfo {author} {\bibfnamefont {X.}~\bibnamefont
  {Sun}}, \bibinfo {author} {\bibfnamefont {H.}~\bibnamefont {Wang}}, \ and\
  \bibinfo {author} {\bibfnamefont {W.~H.}\ \bibnamefont {Miller}},\ }\href
  {\doibase 10.1063/1.477389} {\bibfield  {journal} {\bibinfo  {journal}
  {J.~Chem. Phys.}\ }\textbf {\bibinfo {volume} {109}},\ \bibinfo {pages}
  {7064} (\bibinfo {year} {1998})}\BibitemShut {NoStop}%
\bibitem [{\citenamefont {Wang}\ \emph {et~al.}(1999)\citenamefont {Wang},
  \citenamefont {Song}, \citenamefont {Chandler},\ and\ \citenamefont
  {Miller}}]{Wang1999mapping}%
  \BibitemOpen
  \bibfield  {author} {\bibinfo {author} {\bibfnamefont {H.}~\bibnamefont
  {Wang}}, \bibinfo {author} {\bibfnamefont {X.}~\bibnamefont {Song}}, \bibinfo
  {author} {\bibfnamefont {D.}~\bibnamefont {Chandler}}, \ and\ \bibinfo
  {author} {\bibfnamefont {W.~H.}\ \bibnamefont {Miller}},\ }\href {\doibase
  10.1063/1.478388} {\bibfield  {journal} {\bibinfo  {journal} {J.~Chem.
  Phys.}\ }\textbf {\bibinfo {volume} {110}},\ \bibinfo {pages} {4828}
  (\bibinfo {year} {1999})}\BibitemShut {NoStop}%
\bibitem [{\citenamefont {Cotton}, \citenamefont {Igumenshchev},\ and\
  \citenamefont {Miller}(2014)}]{Cotton2014ET}%
  \BibitemOpen
  \bibfield  {author} {\bibinfo {author} {\bibfnamefont {S.~J.}\ \bibnamefont
  {Cotton}}, \bibinfo {author} {\bibfnamefont {K.}~\bibnamefont
  {Igumenshchev}}, \ and\ \bibinfo {author} {\bibfnamefont {W.~H.}\
  \bibnamefont {Miller}},\ }\href {\doibase 10.1063/1.4893345} {\bibfield
  {journal} {\bibinfo  {journal} {J.~Chem. Phys.}\ }\textbf {\bibinfo {volume}
  {141}},\ \bibinfo {pages} {084104} (\bibinfo {year} {2014})}\BibitemShut
  {NoStop}%
\bibitem [{\citenamefont {Schwerdtfeger}, \citenamefont {Soudackov},\ and\
  \citenamefont {Hammes-Schiffer}(2014)}]{Schwerdtfeger2014ET}%
  \BibitemOpen
  \bibfield  {author} {\bibinfo {author} {\bibfnamefont {C.~A.}\ \bibnamefont
  {Schwerdtfeger}}, \bibinfo {author} {\bibfnamefont {A.~V.}\ \bibnamefont
  {Soudackov}}, \ and\ \bibinfo {author} {\bibfnamefont {S.}~\bibnamefont
  {Hammes-Schiffer}},\ }\href {\doibase 10.1063/1.4855295} {\bibfield
  {journal} {\bibinfo  {journal} {J.~Chem. Phys.}\ }\textbf {\bibinfo {volume}
  {140}},\ \bibinfo {pages} {034113} (\bibinfo {year} {2014})}\BibitemShut
  {NoStop}%
\bibitem [{\citenamefont {Landry}\ and\ \citenamefont
  {Subotnik}(2012)}]{Landry2012hopping}%
  \BibitemOpen
  \bibfield  {author} {\bibinfo {author} {\bibfnamefont {B.~R.}\ \bibnamefont
  {Landry}}\ and\ \bibinfo {author} {\bibfnamefont {J.~E.}\ \bibnamefont
  {Subotnik}},\ }\href {\doibase 10.1063/1.4733675} {\bibfield  {journal}
  {\bibinfo  {journal} {J.~Chem. Phys.}\ }\textbf {\bibinfo {volume} {137}},\
  \bibinfo {pages} {22A513} (\bibinfo {year} {2012})}\BibitemShut {NoStop}%
\bibitem [{\citenamefont {Richardson}\ and\ \citenamefont
  {Thoss}(2013)}]{mapping}%
  \BibitemOpen
  \bibfield  {author} {\bibinfo {author} {\bibfnamefont {J.~O.}\ \bibnamefont
  {Richardson}}\ and\ \bibinfo {author} {\bibfnamefont {M.}~\bibnamefont
  {Thoss}},\ }\href {\doibase 10.1063/1.4816124} {\bibfield  {journal}
  {\bibinfo  {journal} {J.~Chem. Phys.}\ }\textbf {\bibinfo {volume} {139}},\
  \bibinfo {pages} {031102} (\bibinfo {year} {2013})}\BibitemShut {NoStop}%
\bibitem [{\citenamefont {Ananth}(2013)}]{Ananth2013MVRPMD}%
  \BibitemOpen
  \bibfield  {author} {\bibinfo {author} {\bibfnamefont {N.}~\bibnamefont
  {Ananth}},\ }\href {\doibase 10.1063/1.4821590} {\bibfield  {journal}
  {\bibinfo  {journal} {J.~Chem. Phys.}\ }\textbf {\bibinfo {volume} {139}},\
  \bibinfo {pages} {124102} (\bibinfo {year} {2013})}\BibitemShut {NoStop}%
\bibitem [{\citenamefont {Menzeleev}, \citenamefont {Ananth},\ and\
  \citenamefont {Miller}(2011)}]{Menzeleev2011ET}%
  \BibitemOpen
  \bibfield  {author} {\bibinfo {author} {\bibfnamefont {A.~R.}\ \bibnamefont
  {Menzeleev}}, \bibinfo {author} {\bibfnamefont {N.}~\bibnamefont {Ananth}}, \
  and\ \bibinfo {author} {\bibfnamefont {T.~F.}\ \bibnamefont {Miller},
  \bibfnamefont {III}},\ }\href {\doibase 10.1063/1.3624766} {\bibfield
  {journal} {\bibinfo  {journal} {J.~Chem. Phys.}\ }\textbf {\bibinfo {volume}
  {135}},\ \bibinfo {pages} {074106} (\bibinfo {year} {2011})}\BibitemShut
  {NoStop}%
\bibitem [{\citenamefont {Kretchmer}\ and\ \citenamefont
  {Miller~III}(2013)}]{Kretchmer2013ET}%
  \BibitemOpen
  \bibfield  {author} {\bibinfo {author} {\bibfnamefont {J.~S.}\ \bibnamefont
  {Kretchmer}}\ and\ \bibinfo {author} {\bibfnamefont {T.~F.}\ \bibnamefont
  {Miller~III}},\ }\href {\doibase 10.1063/1.4797462} {\bibfield  {journal}
  {\bibinfo  {journal} {J.~Chem. Phys.}\ }\textbf {\bibinfo {volume} {138}},\
  \bibinfo {pages} {134109} (\bibinfo {year} {2013})}\BibitemShut {NoStop}%
\bibitem [{\citenamefont {Menzeleev}, \citenamefont {Bell},\ and\ \citenamefont
  {Miller~III}(2014)}]{Menzeleev2014kinetic}%
  \BibitemOpen
  \bibfield  {author} {\bibinfo {author} {\bibfnamefont {A.~R.}\ \bibnamefont
  {Menzeleev}}, \bibinfo {author} {\bibfnamefont {F.}~\bibnamefont {Bell}}, \
  and\ \bibinfo {author} {\bibfnamefont {T.~F.}\ \bibnamefont {Miller~III}},\
  }\href {\doibase 10.1063/1.4863919} {\bibfield  {journal} {\bibinfo
  {journal} {J.~Chem. Phys.}\ }\textbf {\bibinfo {volume} {140}},\ \bibinfo
  {pages} {064103} (\bibinfo {year} {2014})}\BibitemShut {NoStop}%
\bibitem [{\citenamefont {Shushkov}, \citenamefont {Li},\ and\ \citenamefont
  {Tully}(2012)}]{Shushkov2012RPSH}%
  \BibitemOpen
  \bibfield  {author} {\bibinfo {author} {\bibfnamefont {P.}~\bibnamefont
  {Shushkov}}, \bibinfo {author} {\bibfnamefont {R.}~\bibnamefont {Li}}, \ and\
  \bibinfo {author} {\bibfnamefont {J.~C.}\ \bibnamefont {Tully}},\ }\href
  {\doibase 10.1063/1.4766449} {\bibfield  {journal} {\bibinfo  {journal}
  {J.~Chem. Phys.}\ }\textbf {\bibinfo {volume} {137}},\ \bibinfo {pages}
  {22A549} (\bibinfo {year} {2012})}\BibitemShut {NoStop}%
\bibitem [{\citenamefont {Miller}, \citenamefont {Schwartz},\ and\
  \citenamefont {Tromp}(1983)}]{Miller1983rate}%
  \BibitemOpen
  \bibfield  {author} {\bibinfo {author} {\bibfnamefont {W.~H.}\ \bibnamefont
  {Miller}}, \bibinfo {author} {\bibfnamefont {S.~D.}\ \bibnamefont
  {Schwartz}}, \ and\ \bibinfo {author} {\bibfnamefont {J.~W.}\ \bibnamefont
  {Tromp}},\ }\href {\doibase 10.1063/1.445581} {\bibfield  {journal} {\bibinfo
   {journal} {J.~Chem. Phys.}\ }\textbf {\bibinfo {volume} {79}},\ \bibinfo
  {pages} {4889} (\bibinfo {year} {1983})}\BibitemShut {NoStop}%
\bibitem [{\citenamefont {Richardson}\ and\ \citenamefont
  {Thoss}(2014)}]{nonoscillatory}%
  \BibitemOpen
  \bibfield  {author} {\bibinfo {author} {\bibfnamefont {J.~O.}\ \bibnamefont
  {Richardson}}\ and\ \bibinfo {author} {\bibfnamefont {M.}~\bibnamefont
  {Thoss}},\ }\href {\doibase 10.1063/1.4892865} {\bibfield  {journal}
  {\bibinfo  {journal} {J.~Chem. Phys.}\ }\textbf {\bibinfo {volume} {141}},\
  \bibinfo {pages} {074106} (\bibinfo {year} {2014})},\ \Eprint
  {http://arxiv.org/abs/1406.3144} {arXiv:1406.3144 [physics.chem-ph]}
  \BibitemShut {NoStop}%
\bibitem [{\citenamefont {Wolynes}(1987)}]{Wolynes1987nonadiabatic}%
  \BibitemOpen
  \bibfield  {author} {\bibinfo {author} {\bibfnamefont {P.~G.}\ \bibnamefont
  {Wolynes}},\ }\href {\doibase 10.1063/1.453440} {\bibfield  {journal}
  {\bibinfo  {journal} {J.~Chem. Phys.}\ }\textbf {\bibinfo {volume} {87}},\
  \bibinfo {pages} {6559} (\bibinfo {year} {1987})}\BibitemShut {NoStop}%
\bibitem [{\citenamefont {Bader}, \citenamefont {Kuharski},\ and\ \citenamefont
  {Chandler}(1990)}]{Bader1990golden}%
  \BibitemOpen
  \bibfield  {author} {\bibinfo {author} {\bibfnamefont {J.~S.}\ \bibnamefont
  {Bader}}, \bibinfo {author} {\bibfnamefont {R.~A.}\ \bibnamefont {Kuharski}},
  \ and\ \bibinfo {author} {\bibfnamefont {D.}~\bibnamefont {Chandler}},\
  }\href {\doibase http://dx.doi.org/10.1063/1.459596} {\bibfield  {journal}
  {\bibinfo  {journal} {J.~Chem. Phys.}\ }\textbf {\bibinfo {volume} {93}},\
  \bibinfo {pages} {230} (\bibinfo {year} {1990})}\BibitemShut {NoStop}%
\bibitem [{\citenamefont {Zheng}, \citenamefont {McCammon},\ and\ \citenamefont
  {Wolynes}(1989)}]{Zheng1989ET}%
  \BibitemOpen
  \bibfield  {author} {\bibinfo {author} {\bibfnamefont {C.}~\bibnamefont
  {Zheng}}, \bibinfo {author} {\bibfnamefont {J.~A.}\ \bibnamefont {McCammon}},
  \ and\ \bibinfo {author} {\bibfnamefont {P.~G.}\ \bibnamefont {Wolynes}},\
  }\href {\doibase 10.1073/pnas.86.17.6441} {\bibfield  {journal} {\bibinfo
  {journal} {P. Natl. Acad. Sci. USA}\ }\textbf {\bibinfo {volume} {86}},\
  \bibinfo {pages} {6441} (\bibinfo {year} {1989})}\BibitemShut {NoStop}%
\bibitem [{\citenamefont {Zheng}, \citenamefont {McCammon},\ and\ \citenamefont
  {Wolynes}(1991)}]{Zheng1991ET}%
  \BibitemOpen
  \bibfield  {author} {\bibinfo {author} {\bibfnamefont {C.}~\bibnamefont
  {Zheng}}, \bibinfo {author} {\bibfnamefont {J.~A.}\ \bibnamefont {McCammon}},
  \ and\ \bibinfo {author} {\bibfnamefont {P.~G.}\ \bibnamefont {Wolynes}},\
  }\href {\doibase 10.1016/0301-0104(91)87070-C} {\bibfield  {journal}
  {\bibinfo  {journal} {Chem. Phys.}\ }\textbf {\bibinfo {volume} {158}},\
  \bibinfo {pages} {261} (\bibinfo {year} {1991})}\BibitemShut {NoStop}%
\bibitem [{\citenamefont {Feynman}\ and\ \citenamefont
  {Hibbs}(1965)}]{Feynman}%
  \BibitemOpen
  \bibfield  {author} {\bibinfo {author} {\bibfnamefont {R.~P.}\ \bibnamefont
  {Feynman}}\ and\ \bibinfo {author} {\bibfnamefont {A.~R.}\ \bibnamefont
  {Hibbs}},\ }\href@noop {} {\emph {\bibinfo {title} {Quantum Mechanics and
  Path Integrals}}}\ (\bibinfo  {publisher} {McGraw-Hill},\ \bibinfo {address}
  {New York},\ \bibinfo {year} {1965})\BibitemShut {NoStop}%
\bibitem [{\citenamefont {Miller}(1971)}]{Miller1971density}%
  \BibitemOpen
  \bibfield  {author} {\bibinfo {author} {\bibfnamefont {W.~H.}\ \bibnamefont
  {Miller}},\ }\href {\doibase 10.1063/1.1676560} {\bibfield  {journal}
  {\bibinfo  {journal} {J.~Chem. Phys.}\ }\textbf {\bibinfo {volume} {55}},\
  \bibinfo {pages} {3146} (\bibinfo {year} {1971})}\BibitemShut {NoStop}%
\bibitem [{\citenamefont {Whittaker}(1988)}]{Whittaker}%
  \BibitemOpen
  \bibfield  {author} {\bibinfo {author} {\bibfnamefont {E.~T.}\ \bibnamefont
  {Whittaker}},\ }\href@noop {} {\emph {\bibinfo {title} {A treatise on the
  analytical dynamics of particles and rigid bodies}}}\ (\bibinfo  {publisher}
  {Cambridge},\ \bibinfo {year} {1988})\BibitemShut {NoStop}%
\bibitem [{\citenamefont {Goldstein}, \citenamefont {Poole},\ and\
  \citenamefont {Safko}(2002)}]{Goldstein}%
  \BibitemOpen
  \bibfield  {author} {\bibinfo {author} {\bibfnamefont {H.}~\bibnamefont
  {Goldstein}}, \bibinfo {author} {\bibfnamefont {C.}~\bibnamefont {Poole}}, \
  and\ \bibinfo {author} {\bibfnamefont {J.}~\bibnamefont {Safko}},\
  }\href@noop {} {\emph {\bibinfo {title} {Classical Mechanics}}},\ \bibinfo
  {edition} {3rd}\ ed.\ (\bibinfo  {publisher} {Addison Wesley},\ \bibinfo
  {address} {San Francisco},\ \bibinfo {year} {2002})\BibitemShut {NoStop}%
\bibitem [{\citenamefont {Zhu}\ \emph {et~al.}(1997)\citenamefont {Zhu},
  \citenamefont {Byrd}, \citenamefont {Lu},\ and\ \citenamefont
  {Nocedal}}]{LBFGSB1997algorithm}%
  \BibitemOpen
  \bibfield  {author} {\bibinfo {author} {\bibfnamefont {C.}~\bibnamefont
  {Zhu}}, \bibinfo {author} {\bibfnamefont {R.~H.}\ \bibnamefont {Byrd}},
  \bibinfo {author} {\bibfnamefont {P.}~\bibnamefont {Lu}}, \ and\ \bibinfo
  {author} {\bibfnamefont {J.}~\bibnamefont {Nocedal}},\ }\href {\doibase
  http://doi.acm.org/10.1145/279232.279236} {\bibfield  {journal} {\bibinfo
  {journal} {ACM Trans. Math. Softw.}\ }\textbf {\bibinfo {volume} {23}},\
  \bibinfo {pages} {550} (\bibinfo {year} {1997})}\BibitemShut {NoStop}%
\bibitem [{\citenamefont {Press}\ \emph {et~al.}(1992)\citenamefont {Press},
  \citenamefont {Teukolsky}, \citenamefont {Vetterling},\ and\ \citenamefont
  {Flannery}}]{NumRep}%
  \BibitemOpen
  \bibfield  {author} {\bibinfo {author} {\bibfnamefont {W.~H.}\ \bibnamefont
  {Press}}, \bibinfo {author} {\bibfnamefont {S.~A.}\ \bibnamefont
  {Teukolsky}}, \bibinfo {author} {\bibfnamefont {W.~T.}\ \bibnamefont
  {Vetterling}}, \ and\ \bibinfo {author} {\bibfnamefont {B.~P.}\ \bibnamefont
  {Flannery}},\ }\href@noop {} {\emph {\bibinfo {title} {Numerical Recipes in
  Fortran 77: {The} Art of Scientific Computing}}},\ \bibinfo {edition} {2nd}\
  ed.\ (\bibinfo  {publisher} {Cambridge University Press},\ \bibinfo {address}
  {Cambridge},\ \bibinfo {year} {1992})\BibitemShut {NoStop}%
\bibitem [{\citenamefont {Parrinello}\ and\ \citenamefont
  {Rahman}(1984)}]{Parrinello1984Fcenter}%
  \BibitemOpen
  \bibfield  {author} {\bibinfo {author} {\bibfnamefont {M.}~\bibnamefont
  {Parrinello}}\ and\ \bibinfo {author} {\bibfnamefont {A.}~\bibnamefont
  {Rahman}},\ }\href {\doibase 10.1063/1.446740} {\bibfield  {journal}
  {\bibinfo  {journal} {J.~Chem. Phys.}\ }\textbf {\bibinfo {volume} {80}},\
  \bibinfo {pages} {860} (\bibinfo {year} {1984})}\BibitemShut {NoStop}%
\bibitem [{\citenamefont {Habershon}\ \emph {et~al.}(2013)\citenamefont
  {Habershon}, \citenamefont {Manolopoulos}, \citenamefont {Markland},\ and\
  \citenamefont {Miller~III}}]{Habershon2013RPMDreview}%
  \BibitemOpen
  \bibfield  {author} {\bibinfo {author} {\bibfnamefont {S.}~\bibnamefont
  {Habershon}}, \bibinfo {author} {\bibfnamefont {D.~E.}\ \bibnamefont
  {Manolopoulos}}, \bibinfo {author} {\bibfnamefont {T.~E.}\ \bibnamefont
  {Markland}}, \ and\ \bibinfo {author} {\bibfnamefont {T.~F.}\ \bibnamefont
  {Miller~III}},\ }\href {\doibase 10.1146/annurev-physchem-040412-110122}
  {\bibfield  {journal} {\bibinfo  {journal} {Annu. Rev. Phys. Chem.}\ }\textbf
  {\bibinfo {volume} {64}},\ \bibinfo {pages} {387} (\bibinfo {year}
  {2013})}\BibitemShut {NoStop}%
\bibitem [{\citenamefont {Chandler}\ and\ \citenamefont
  {Wolynes}(1981)}]{Chandler+Wolynes1981}%
  \BibitemOpen
  \bibfield  {author} {\bibinfo {author} {\bibfnamefont {D.}~\bibnamefont
  {Chandler}}\ and\ \bibinfo {author} {\bibfnamefont {P.~G.}\ \bibnamefont
  {Wolynes}},\ }\href {\doibase 10.1063/1.441588} {\bibfield  {journal}
  {\bibinfo  {journal} {J.~Chem. Phys.}\ }\textbf {\bibinfo {volume} {74}},\
  \bibinfo {pages} {4078} (\bibinfo {year} {1981})}\BibitemShut {NoStop}%
\bibitem [{\citenamefont {Kleinert}(2006)}]{Kleinert}%
  \BibitemOpen
  \bibfield  {author} {\bibinfo {author} {\bibfnamefont {H.}~\bibnamefont
  {Kleinert}},\ }\href@noop {} {\emph {\bibinfo {title} {Path Integrals in
  Quantum Mechanics, Statistics, Polymer Physics and Financial Markets}}},\
  \bibinfo {edition} {4th}\ ed.\ (\bibinfo  {publisher} {World Scientific},\
  \bibinfo {address} {Singapore},\ \bibinfo {year} {2006})\BibitemShut
  {NoStop}%
\bibitem [{\citenamefont {Cao}\ and\ \citenamefont
  {Voth}(1997)}]{Cao1997nonadiabatic}%
  \BibitemOpen
  \bibfield  {author} {\bibinfo {author} {\bibfnamefont {J.}~\bibnamefont
  {Cao}}\ and\ \bibinfo {author} {\bibfnamefont {G.~A.}\ \bibnamefont {Voth}},\
  }\href {\doibase 10.1063/1.474123} {\bibfield  {journal} {\bibinfo  {journal}
  {J.~Chem. Phys.}\ }\textbf {\bibinfo {volume} {106}},\ \bibinfo {pages}
  {1769} (\bibinfo {year} {1997})}\BibitemShut {NoStop}%
\bibitem [{\citenamefont {Cao}\ and\ \citenamefont
  {Voth}(1998)}]{Cao1998erratum}%
  \BibitemOpen
  \bibfield  {author} {\bibinfo {author} {\bibfnamefont {J.}~\bibnamefont
  {Cao}}\ and\ \bibinfo {author} {\bibfnamefont {G.~A.}\ \bibnamefont {Voth}},\
  }\href {\doibase 10.1063/1.476782} {\bibfield  {journal} {\bibinfo  {journal}
  {J.~Chem. Phys.}\ }\textbf {\bibinfo {volume} {109}},\ \bibinfo {pages}
  {2043} (\bibinfo {year} {1998})}\BibitemShut {NoStop}%
\bibitem [{\citenamefont {Van{\'\i}{\v{c}}ek}\ \emph
  {et~al.}(2005)\citenamefont {Van{\'\i}{\v{c}}ek}, \citenamefont {Miller},
  \citenamefont {Castillo},\ and\ \citenamefont {Aoiz}}]{Vanicek2005QI}%
  \BibitemOpen
  \bibfield  {author} {\bibinfo {author} {\bibfnamefont {J.}~\bibnamefont
  {Van{\'\i}{\v{c}}ek}}, \bibinfo {author} {\bibfnamefont {W.~H.}\ \bibnamefont
  {Miller}}, \bibinfo {author} {\bibfnamefont {J.~F.}\ \bibnamefont
  {Castillo}}, \ and\ \bibinfo {author} {\bibfnamefont {F.~J.}\ \bibnamefont
  {Aoiz}},\ }\href {\doibase 10.1063/1.1946740} {\bibfield  {journal} {\bibinfo
   {journal} {J.~Chem. Phys.}\ }\textbf {\bibinfo {volume} {123}},\ \bibinfo
  {pages} {054108} (\bibinfo {year} {2005})}\BibitemShut {NoStop}%
\bibitem [{\citenamefont {Miller}\ \emph {et~al.}(2003)\citenamefont {Miller},
  \citenamefont {Zhao}, \citenamefont {Ceotto},\ and\ \citenamefont
  {Yang}}]{Miller2003QI}%
  \BibitemOpen
  \bibfield  {author} {\bibinfo {author} {\bibfnamefont {W.~H.}\ \bibnamefont
  {Miller}}, \bibinfo {author} {\bibfnamefont {Y.}~\bibnamefont {Zhao}},
  \bibinfo {author} {\bibfnamefont {M.}~\bibnamefont {Ceotto}}, \ and\ \bibinfo
  {author} {\bibfnamefont {S.}~\bibnamefont {Yang}},\ }\href {\doibase
  10.1063/1.1580110} {\bibfield  {journal} {\bibinfo  {journal} {J.~Chem.
  Phys.}\ }\textbf {\bibinfo {volume} {119}},\ \bibinfo {pages} {1329}
  (\bibinfo {year} {2003})}\BibitemShut {NoStop}%
\bibitem [{\citenamefont {Weiss}(2008)}]{Weiss}%
  \BibitemOpen
  \bibfield  {author} {\bibinfo {author} {\bibfnamefont {U.}~\bibnamefont
  {Weiss}},\ }\href@noop {} {\emph {\bibinfo {title} {Quantum Dissipative
  Systems}}},\ \bibinfo {edition} {3rd}\ ed.\ (\bibinfo  {publisher} {World
  Scientific},\ \bibinfo {address} {Singapore},\ \bibinfo {year}
  {2008})\BibitemShut {NoStop}%
\bibitem [{\citenamefont {Marcus}\ and\ \citenamefont
  {Sutin}(1985)}]{Marcus1985review}%
  \BibitemOpen
  \bibfield  {author} {\bibinfo {author} {\bibfnamefont {R.~A.}\ \bibnamefont
  {Marcus}}\ and\ \bibinfo {author} {\bibfnamefont {N.}~\bibnamefont {Sutin}},\
  }\href {\doibase 10.1016/0304-4173(85)90014-X} {\bibfield  {journal}
  {\bibinfo  {journal} {Biochim. Biophys. Acta}\ }\textbf {\bibinfo {volume}
  {811}},\ \bibinfo {pages} {265} (\bibinfo {year} {1985})}\BibitemShut
  {NoStop}%
\bibitem [{\citenamefont {Richardson}(2012)}]{PhD}%
  \BibitemOpen
  \bibfield  {author} {\bibinfo {author} {\bibfnamefont {J.~O.}\ \bibnamefont
  {Richardson}},\ }\emph {\bibinfo {title} {Ring-Polymer Approaches to
  Instanton Theory}},\ \href {www.dspace.cam.ac.uk/handle/1810/243641} {Ph.D.
  thesis},\ \bibinfo  {school} {University of Cambridge} (\bibinfo {year}
  {2012})\BibitemShut {NoStop}%
\bibitem [{\citenamefont {Faccioli}\ \emph {et~al.}(2006)\citenamefont
  {Faccioli}, \citenamefont {Sega}, \citenamefont {Pederiva},\ and\
  \citenamefont {Orland}}]{Faccioli2006pathways}%
  \BibitemOpen
  \bibfield  {author} {\bibinfo {author} {\bibfnamefont {P.}~\bibnamefont
  {Faccioli}}, \bibinfo {author} {\bibfnamefont {M.}~\bibnamefont {Sega}},
  \bibinfo {author} {\bibfnamefont {F.}~\bibnamefont {Pederiva}}, \ and\
  \bibinfo {author} {\bibfnamefont {H.}~\bibnamefont {Orland}},\ }\href
  {\doibase 10.1103/PhysRevLett.97.108101} {\bibfield  {journal} {\bibinfo
  {journal} {Phys. Rev. Lett.}\ }\textbf {\bibinfo {volume} {97}},\ \bibinfo
  {pages} {108101} (\bibinfo {year} {2006})}\BibitemShut {NoStop}%
\bibitem [{\citenamefont {a~Beccara}\ \emph {et~al.}(2012)\citenamefont
  {a~Beccara}, \citenamefont {{\v{S}}krbi{\'c}}, \citenamefont {Covino},\ and\
  \citenamefont {Faccioli}}]{Beccara2012folding}%
  \BibitemOpen
  \bibfield  {author} {\bibinfo {author} {\bibfnamefont {S.}~\bibnamefont
  {a~Beccara}}, \bibinfo {author} {\bibfnamefont {T.}~\bibnamefont
  {{\v{S}}krbi{\'c}}}, \bibinfo {author} {\bibfnamefont {R.}~\bibnamefont
  {Covino}}, \ and\ \bibinfo {author} {\bibfnamefont {P.}~\bibnamefont
  {Faccioli}},\ }\href {\doibase 10.1073/pnas.1111796109} {\bibfield  {journal}
  {\bibinfo  {journal} {P. Natl. Acad. Sci. USA.}\ }\textbf {\bibinfo {volume}
  {109}},\ \bibinfo {pages} {2330} (\bibinfo {year} {2012})}\BibitemShut
  {NoStop}%
\bibitem [{\citenamefont {Einarsd{\'o}ttir}\ \emph {et~al.}(2012)\citenamefont
  {Einarsd{\'o}ttir}, \citenamefont {Arnaldsson}, \citenamefont
  {{\'O}skarsson},\ and\ \citenamefont {J{\'o}nsson}}]{Einarsdottir2012path}%
  \BibitemOpen
  \bibfield  {author} {\bibinfo {author} {\bibfnamefont {D.~M.}\ \bibnamefont
  {Einarsd{\'o}ttir}}, \bibinfo {author} {\bibfnamefont {A.}~\bibnamefont
  {Arnaldsson}}, \bibinfo {author} {\bibfnamefont {F.}~\bibnamefont
  {{\'O}skarsson}}, \ and\ \bibinfo {author} {\bibfnamefont {H.}~\bibnamefont
  {J{\'o}nsson}},\ }in\ \href@noop {} {\emph {\bibinfo {booktitle} {Applied
  Parallel and Scientific Computing}}},\ \bibinfo {series} {Lecture Notes in
  Computer Science}, Vol.\ \bibinfo {volume} {7134}\ (\bibinfo  {publisher}
  {Springer},\ \bibinfo {year} {2012})\ pp.\ \bibinfo {pages}
  {45--55}\BibitemShut {NoStop}%
\bibitem [{\citenamefont {Garg}, \citenamefont {Onuchic},\ and\ \citenamefont
  {Ambegaokar}(1985)}]{Garg1985spinboson}%
  \BibitemOpen
  \bibfield  {author} {\bibinfo {author} {\bibfnamefont {A.}~\bibnamefont
  {Garg}}, \bibinfo {author} {\bibfnamefont {J.~N.}\ \bibnamefont {Onuchic}}, \
  and\ \bibinfo {author} {\bibfnamefont {V.}~\bibnamefont {Ambegaokar}},\
  }\href {\doibase 10.1063/1.449017} {\bibfield  {journal} {\bibinfo  {journal}
  {J.~Chem. Phys.}\ }\textbf {\bibinfo {volume} {83}},\ \bibinfo {pages} {4491}
  (\bibinfo {year} {1985})}\BibitemShut {NoStop}%
\bibitem [{\citenamefont {Leggett}\ \emph {et~al.}(1987)\citenamefont
  {Leggett}, \citenamefont {Chakravarty}, \citenamefont {Dorsey}, \citenamefont
  {Fisher}, \citenamefont {Garg},\ and\ \citenamefont
  {Zwerger}}]{Leggett1987spinboson}%
  \BibitemOpen
  \bibfield  {author} {\bibinfo {author} {\bibfnamefont {A.~J.}\ \bibnamefont
  {Leggett}}, \bibinfo {author} {\bibfnamefont {S.}~\bibnamefont
  {Chakravarty}}, \bibinfo {author} {\bibfnamefont {A.~T.}\ \bibnamefont
  {Dorsey}}, \bibinfo {author} {\bibfnamefont {M.~P.~A.}\ \bibnamefont
  {Fisher}}, \bibinfo {author} {\bibfnamefont {A.}~\bibnamefont {Garg}}, \ and\
  \bibinfo {author} {\bibfnamefont {W.}~\bibnamefont {Zwerger}},\ }\href
  {\doibase 10.1103/RevModPhys.59.1} {\bibfield  {journal} {\bibinfo  {journal}
  {Rev. Mod. Phys.}\ }\textbf {\bibinfo {volume} {59}},\ \bibinfo {pages} {1}
  (\bibinfo {year} {1987})}\BibitemShut {NoStop}%
\bibitem [{\citenamefont {Wang}\ and\ \citenamefont
  {Thoss}(2003)}]{Wang2003RuRu}%
  \BibitemOpen
  \bibfield  {author} {\bibinfo {author} {\bibfnamefont {H.}~\bibnamefont
  {Wang}}\ and\ \bibinfo {author} {\bibfnamefont {M.}~\bibnamefont {Thoss}},\
  }\href {\doibase 10.1021/jp0272668} {\bibfield  {journal} {\bibinfo
  {journal} {J.~Phys.\ Chem.~A}\ }\textbf {\bibinfo {volume} {107}},\ \bibinfo
  {pages} {2126} (\bibinfo {year} {2003})}\BibitemShut {NoStop}%
\bibitem [{\citenamefont {Berkelbach}, \citenamefont {Reichman},\ and\
  \citenamefont {Markland}(2012)}]{Berkelbach2012hybrid}%
  \BibitemOpen
  \bibfield  {author} {\bibinfo {author} {\bibfnamefont {T.~C.}\ \bibnamefont
  {Berkelbach}}, \bibinfo {author} {\bibfnamefont {D.~R.}\ \bibnamefont
  {Reichman}}, \ and\ \bibinfo {author} {\bibfnamefont {T.~E.}\ \bibnamefont
  {Markland}},\ }\href {\doibase 10.1063/1.3671372} {\bibfield  {journal}
  {\bibinfo  {journal} {J.~Chem. Phys.}\ }\textbf {\bibinfo {volume} {136}},\
  \bibinfo {pages} {034113} (\bibinfo {year} {2012})}\BibitemShut {NoStop}%
\bibitem [{\citenamefont {Esquinazi}(1998)}]{EsquinaziBook}%
  \BibitemOpen
  \bibinfo {editor} {\bibfnamefont {P.}~\bibnamefont {Esquinazi}},\ ed.,\
  \href@noop {} {\emph {\bibinfo {title} {Tunneling Systems in Amorphous and
  Crystalline Solids}}}\ (\bibinfo  {publisher} {Springer},\ \bibinfo {year}
  {1998})\BibitemShut {NoStop}%
\end{thebibliography}%

\end{document}